\documentclass[12pt, draftclsnofoot, onecolumn]{IEEEtran}
\usepackage{graphicx}
\usepackage{epstopdf}
\usepackage{mathrsfs}
\usepackage{amsmath}
\usepackage{latexsym}
\usepackage{multirow}
\usepackage{caption}
\usepackage{booktabs}
\usepackage{xcolor}
\usepackage{color}
\DeclareCaptionFont{red}{\color{red}}
\DeclareCaptionFont{green}{\color{green}}
\DeclareCaptionFont{blue}{\color{blue}}
\usepackage{subfigure}
\usepackage{setspace}
\usepackage{algorithm, algorithmic}
\usepackage{bbm}
\usepackage{url}
\usepackage{color}
\usepackage[amsmath,thmmarks]{ntheorem}
\usepackage{bm}
\usepackage{caption}
\def\IEEEproofindentspace{0\parindent}
\newtheorem{theorem}{Theorem}

\newtheorem{corollary}{Corollary}
\graphicspath{{lunwen/}}
\allowdisplaybreaks[4]

\begin{document}

\title{\huge A Machine Learning Approach for Task and Resource Allocation in Mobile Edge Computing Based Networks}
\vspace{-0.2cm}

\author{\large{Sihua Wang,} \emph{Student Member}, \emph{IEEE}, {Mingzhe Chen,} \emph{Member}, \emph{IEEE},\\ {Xuanlin Liu,} \emph{Student Member}, \emph{IEEE}, {Changchuan Yin}, \emph{Senior Member}, \emph{IEEE},\\ {Shuguang Cui}, \emph{Fellow}, \emph{IEEE}, and {H. Vincent Poor}, \emph{Fellow}, \emph{IEEE}\\
\thanks{S. Wang, X. Liu, and C. Yin are with the Beijing Laboratory of Advanced Information Network, and the Beijing Key Laboratory of Network System Architecture and Convergence, Beijing University of Posts and Telecommunications, Beijing 100876, China. Email: \protect\url{sihuawang@bupt.edu.cn;} xuanlin.liu@bupt.edu.cn; ccyin@ieee.org.}
\thanks{M. Chen is  with the Department of Electrical Engineering, Princeton University, Princeton, NJ, 08544, USA, and also with the Chinese University of Hong Kong, Shenzhen, 518172, China, Email: \protect\url{mingzhec@princeton.edu}.}
\thanks{S. Cui is with the Shenzhen Research Institute of Big Data and Future Network of Intelligence Institute (FNii), the Chinese University of Hong Kong, Shenzhen, 518172, China, Email: \protect\url{shuguangcui@cuhk.edu.cn}.}
\thanks{H. V. Poor is with the Department of Electrical Engineering, Princeton University, Princeton, NJ, 08544, USA, Email: \protect\url{poor@princeton.edu}.}
}

\maketitle
\vspace{-1.5cm}
\begin{abstract}
In this paper, a joint task, spectrum, and transmit power allocation problem is investigated for a wireless network in which the base stations (BSs) are equipped with mobile edge computing (MEC) servers to jointly provide computational and communication services to users. Each user can request one computational task from three types of computational tasks. Since the data size of each computational task is different, as the requested computational task varies, the BSs must adjust their resource (subcarrier and transmit power) and task allocation schemes to effectively serve the users. This problem is formulated as an optimization problem whose goal is to minimize the maximal computational and transmission delay among all users. A multi-stack reinforcement learning (RL) algorithm is developed to solve this problem. Using the proposed algorithm, each BS can record the historical resource allocation schemes and users' information in its multiple stacks to avoid learning the same resource allocation scheme and users' states, thus improving the convergence speed and learning efficiency. Simulation results illustrate that the proposed algorithm can reduce the number of iterations needed for convergence and the maximal delay among all users by up to 18\% and 11.1\% compared to the standard Q-learning algorithm.
\end{abstract}

\vspace{-0.1cm}

\begin{IEEEkeywords}
Mobile edge computing, resource management, multi-stack reinforcement learning.
\end{IEEEkeywords}
\IEEEpeerreviewmaketitle

\section{Introduction}

Since the multimedia and real-time applications such as augmented reality require powerful computational capability \cite{PZ}, mobile devices with limited computational capability may not be able to perform these novel applications \cite{FRZY}. To overcome this issue, mobile edge computing (MEC) servers can be deployed at the wireless base stations (BSs) to help mobile devices process their computational tasks \cite{ASWD}. However, the deployment of MEC servers over wireless networks also faces a number of challenges such as the optimization of MEC server deployment, task allocation, and energy efficiency \cite{FGMY}.

A number of existing works studied important problems related to wireless and computational resource allocation such as in \cite{XLW}--[13]. In \cite{XLW}, the authors maximized the spectrum efficiency via optimizing computational task allocation. The authors in \cite{ZJM} studied the minimization of the energy consumption of all users using MEC. In \cite{LS}, the authors optimized the energy efficiency of each user in MEC based networks. However, the existing works \cite{XLW}--[7] only optimized the resource allocation for one BS. Hence, they may not be suitable for a network with several BSs. The authors in \cite{ZCKM} studied the multi-user computational task offloading problem to minimize the users' energy consumption. In \cite{SY}, the authors proposed a binary computational task offloading scheme to maximize the total throughput. The work in \cite{JLS} developed a resource management algorithm to minimize the long-term system energy cost. The authors in \cite{CKHB} developed a task offloading scheme to minimize the energy consumption. However, the existing works in \cite{ZCKM}--[11] that studied the resource allocation policies assuming that all users request a computational task that can be offloaded to the MEC servers, did not consider the scenario in which the types of requested computational tasks are different (e.g., some users must process the computational task locally and other users can process the computational tasks with the help of MEC servers). The computational and communication resources required for processing different types of tasks are different \cite{XFJ}. For example, a task that is performed by both the user and the MEC server needs more communication resource than a task that is processed by user itself \cite{YMZT}. Meanwhile, as the data size of each computational task requested by each user varies, the BSs need to rerun their optimization algorithms to cope with this change thus resulting in additional overhead and delay for computational task processing \cite{ZJDP}. To solve this problem, one promising solution is to use reinforcement learning (RL) approach since RL algorithms can find a relationship between the users' computational tasks and the resource allocation policy so as to directly generate the resource allocation policy without the time consumption for finding the optimal resource allocation strategy \cite{YFMZ}.

The existing literature in \cite{TND}--\cite{RCWYB} studied the use of RL algorithms for solving MEC related problems. The work in \cite{TND} developed a federated RL to minimize the sum of the energy consumption of the devices. In \cite{XCX}, the authors used a federated deep RL approach to optimize the caching strategy in an MEC-based network. However, the works in \cite{TND} and \cite{XCX} require the BSs to exchange the resource allocation scheme and each user's state, thus increasing communication overhead. In \cite{YFYR}, the authors proposed a model-free RL task offloading mechanism to minimize the energy consumption of users. An RL algorithm is used in \cite{RCWYB} to maximize the throughput of the BSs under the constraint of communication cost of each user. However, the RL algorithms in most of these existing works \cite{YFYR}--\cite{RCWYB} may repeatedly learn the same resource allocation scheme during the training process thus increase RL convergence time. Therefore, it is necessary to develop a novel algorithm that can avoid learning the same resource allocation scheme and improve the learning efficiency.

The main contribution of this paper is a novel resource allocation framework for an MEC-based network with the users who can request different computational tasks. In summary, the main contributions of the paper are:
\begin{itemize}
\item We consider an MEC-based network in which each user can request different computational tasks. Different from the existing works that consider only a single type of computation tasks [8]--[11], we assume that each user can request different types of computational tasks. To effectively serve the users, a novel resource allocation scheme must be developed. This problem is formulated as an optimization problem aiming to minimize the maximum computation and transmission delay among all users.

\item To solve the proposed problem, we develop a multi-stack RL method. Compared to the conventional RL algorithms in \cite{TND}--\cite{RCWYB}, the proposed algorithm uses multiple stacks to record historical resource allocation schemes and users' states,
    which can avoid learning the same information, thus improving the convergence speed and the learning efficiency.

\item We perform fundamental analysis on the gains that stem from the change of the transmit power and the subcarriers over uplink and downlink for each user. The analytical result shows that, to reduce the maximum delay among all users, each BS prefers to allocate more downlink subcarriers and the downlink transmit power to a user with a task that must be processed by the MEC server. In contrast, each BS prefers to allocate more uplink subcarriers and the uplink transmit power to a user with a task that must be locally processed.

\end{itemize}

Simulation results illustrate that the proposed RL algorithm can reduce the number of iterations needed for convergence and the maximal delay among all users by up to 18\% and 11.1\% compared to Q-learning. To the best of our knowledge, this is the first work that \emph{studies the use of multi-stack RL method to optimize the resource allocation in an MEC  based network}.

The rest of this paper is organized as follows. The system model and the problem formulation are described in Section II. The multiple stack RL method for resource and task allocation is presented in Section III. In Section IV, numerical results are presented and discussed. Finally, conclusions are drawn in Section V.

\vspace{-0.1cm}
\section{System Model and Problem Formulation}
\vspace{-0.1cm}
\begin{figure}
\centering
\setlength{\belowcaptionskip}{-0.7cm}
\includegraphics[width=15cm]{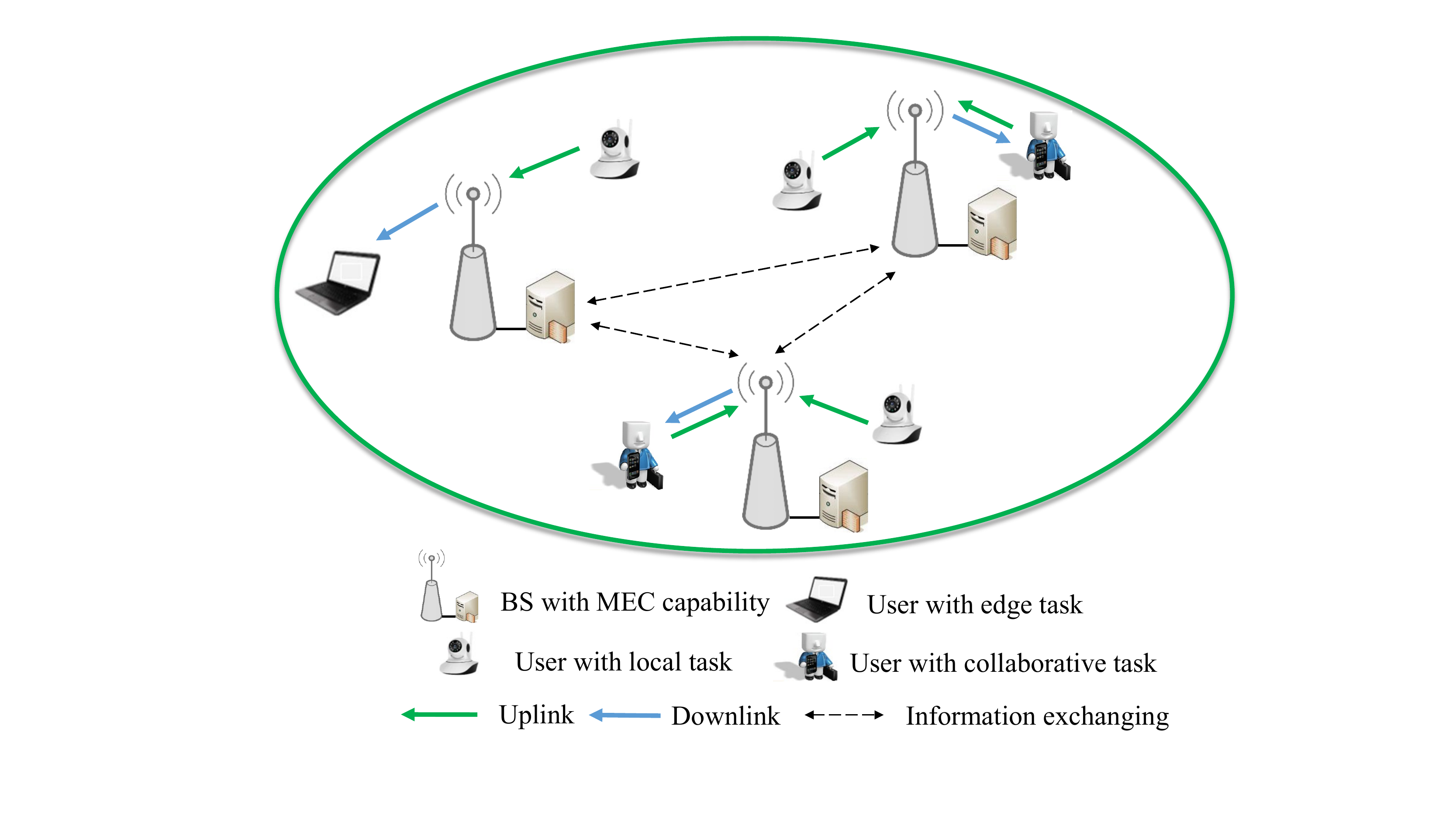}
\vspace{-0.5cm}
\caption*{Fig. 1. The architecture of an MEC based network.}
\label{3}
\end{figure}

We consider an MEC-based network with a set $\mathcal{N}$ of $N$ BSs serving a set of $\mathcal{M}$ of $M$ users, as shown in Fig. 1. In our model, each user can only connect to one BS for task processing and each BS can simultaneously execute multiple computational tasks requested by its associated users \cite{YFM}.

\subsection{Transmission Model}
The orthogonal frequency division multiple access (OFDMA) transmission scheme is adopt for each BS \cite{YWG}. Let $\mathcal{I}\!=\!\{1,2,\ldots,I\}$ and $\mathcal{J}\!=\!\{1,2,\ldots,J\}$ be the set of uplink orthogonal subcarriers and downlink orthogonal subcarriers, respectively.  Given a bandwidth $W$ for each uplink or downlink subcarrier, the uplink and downlink data rates of user $m$ associated with BS $n$ over uplink subcarrier $i \in \mathcal{I}$ and downlink subcarrier $j \in \mathcal{J}$ can be given by (in bits/s) \cite{GWJ}:
\vspace{-0.1cm}
\begin{equation}\label{4}
U_{n,m}^{i}\!\!\left({v_{n,m}^i},\!{u_{n,m}^i}\right)\!\!=\! {u_{n,m}^i}W{\log _2}\!\left(\!\! {1\!+\!\frac{{v_{n,m}^i{\left|h_{n,m}^i\right|}^2}}{\sigma^2_{N}\!+\!\!\!\! \sum\limits_{\substack{p \in\mathcal{M},\\ p\ne m}}\! \!\!{u_{n,p}^i}{v_{n,p}^{i}{\left|h_{n,p}^i\right|^2}}  }}\!\right)\!\!,
\end{equation}
\begin{equation}\label{4}
\!D_{n,m}^j\!\!\left({w_{n,m}^j},{d_{n,m}^j} \right)\!\!=\!{d_{n,m}^j}W{\log _2}\!\!\left( \!\!{1\! +\! \frac{{w_{n,m}^j}{\left|h_{n,m}^j\right|}^2}{\sigma^2_{N}\!\!+\!\!\!\!\sum\limits_{\substack{p \in \mathcal{N},\\ p\ne n}}\!\!{{d_{p,m}^j}{w_{p,m}^j}{\left|h_{p,m}^j\right|}^2} \!\!\!} } \right)\!\!,
\end{equation}
respectively, where ${v_{n,m}^i}$ is user $m$'s transmit power on uplink subcarrier $i$ and ${w_{n,m}^j}$ is BS $n$'s transmit power on downlink subcarrier $j$. $h_{n,m}^i = g_{n,m}^ir^{-\delta}_{n,m}$ and $h_{n,m}^j \!=\! g_{n,m}^jr^{-\delta}_{n,m}$ is the channel gain between user $m$ and BS $n$ over subcarrier $i$ and $j$, respectively. Here, $g_{n,m}^i$ and $g_{n,m}^j$ are the Rayleigh fading parameters, $r_{n,m}$ is the distance between user $m$ and BS $n$, and $\delta$ is the path loss exponent. $\sigma^2_{N}$ is the power of the Gaussian noise. $u_{n,m}^i$ is the uplink subcarrier allocation index with $u_{n,m}^i\!=\!1$ indicating that user $m$ associates with BS \emph{n} using subcarrier $i$, and otherwise, we have $u_{n,m}^i=0$. ${d_{n,m}^j}$ is the downlink subcarrier allocation index with ${d_{n,m}^j}=1$ indicating that BS $n$ connects to user $m$ using subcarrier $j$, and ${d_{n,m}^j}=0$, otherwise.

The sum transmission rate over uplink and downlink between user \emph{m} and BS \emph{n} is:
\begin{equation}\label{4}
\vspace{-0.1cm}
{U_{n,m}}\!\left( {\bm v_{n,m},{\bm u_{n,m}}} \right)\!=\!\! \sum\limits_{i \in \cal{I}} {U_{n,m}^i} \!\left( {v_{n,m}^i},\!{u_{n,m}^i} \right),\,
 \end{equation}
\begin{equation}\label{4}
  \setlength{\abovedisplayshortskip}{-0.1cm}
{D_{n,m}}\!\left( {\bm w_{n,m},{\bm d_{n,m}}} \right)\!=\!\!\! \sum\limits_{j \in \cal{J}} \!{D_{n,m}^j} \!\left( {w_{n,m}^j},\!{d_{n,m}^j} \right),
 \end{equation}
\vspace{-0.5cm}
 \begin{flushleft}where $\bm v_{n,m}\!\!=\!\![v_{n,m}^{1},\!\ldots\!,\!v_{n,m}^{I}]$, $\bm w_{n,m}\!\!=\!\![w_{n,m}^{1},\!\ldots\!,w_{n,m}^{J}]$, $\bm u_{n,m}\!\!=\!\![u_{n,m}^{1},\!\ldots\!,u_{n,m}^I]$, and $\bm d_{n,m}\!\!=\!\![d_{n,m}^{1},\!\ldots\!,$\\$d_{n,m}^J]$.
 \end{flushleft}
 \vspace{-0.1cm}
\subsection{Computation Model}
\vspace{-0.1cm}
We assume that each user can request one computational task from three types of computational tasks, specified as follows:

\begin{itemize}
  \item \textbf{Edge task}: Edge tasks requested by users must be completely computed by MEC servers \cite{WSS}. Then, the computational result must be transmitted to the users. For example, when a user wants to watch a movie on the mobile device, the BS must compress the video before this video is transmitted to the user \cite{LXCX}. The time used to process an edge task requested by user $m$ is given by:
      \vspace{-0.1cm}
  \begin{equation}\label{6}
  \begin{aligned}
   {t^1_m\left( {\bm{w}_{n,m}},{\bm{d}_{n,m}} \right)}
   = \frac{{\omega {\lambda _m}}}{F_{}} + \frac{\nu\lambda _m}{{D_{n,m}}\left( {\bm{w}_{n,m}},{\bm{d}_{n,m}} \right)},\
   \end{aligned}
  \end{equation}where \emph{F} is the CPU clock frequency of an MEC server. $\omega$ represents the number of CPU cycles used to compute one bit data at an MEC server. $\lambda_\emph{m}$ is the data size of the computational task of user \emph{m}. $\nu$ is a constant to represent the ratio between the data size of each computational task before processing and the data size of the computational result after processing. \emph{F}, $\omega$, and $\nu$ are assumed to be equal for all MEC servers. The first term represents the time consumption for computing the task requested by user $m$ in the MEC server and the second term represents the time consumption for transmitting the computational result to user \emph{m}.

\vspace{0.1cm}
  \item \textbf{Local task}: A local task must be completely computed at mobile devices and then transmitted to the BS \cite{YMX}. For example, when a user wants to upload photos to Twitter, the user must compress the images locally before they are transmitted to the BS \cite{MWC}. The time that user $m$ uses to compute its local task is given by:
  \begin{equation}\label{5}
  \begin{aligned}
   {t^2_m\left( {\bm{v}_{n,m}},{\bm{u}_{n,m}} \right)} = \frac{{{\omega _m}{\lambda _m}}}{{{f_m}}} \!+\! \frac{\nu\lambda _m}{{U_{n,m}}\left( {\bm{v}_{n,m}},{\bm{u}_{n,m}} \right)},\
    \end{aligned}
  \end{equation}where \emph{f}$_m$ is the CPU clock frequency of each user $m$ and $\omega_\emph{m}$ is the number of CPU cycles used to compute one bit data at each user \emph{m}. The first term implies the time consumption for computing the task locally and the second term implies the time consumption for transmitting the computational result to BS \emph{n}.

\vspace{0.1cm}
  \item \textbf{Collaborative task}: Each collaborative task can be divided into a local computational task processed by a user and and an edge computational task processed by an MEC server \cite{YFS}. For example, when a user plays a virtual reality (VR) online games, the BS must collect the tracking information from the user and then, transmit the generated VR image to the user \cite{WGM}. The time consumption for processing the collaborative task can be given by:
   \begin{equation}
   \begin{array}{l}
   \begin{aligned}
   &{t^3_m}\left( {\bm{v}_{n,m},{\bm{u}_{n,m}},{\bm{w}_{n,m}},{\bm{d}_{n,m}},{\mu _m}} \right)\\
   &{\rm{ = }}\max \left\{ {\frac{{\omega _m}{\mu _m}{\lambda _m}}{f_m},} \right.\frac{{(1 - {\mu _m}){\lambda _m}}}{{{U_{m,n}}\left( {\bm{v}_{n,m},{\bm{u}_{n,m}}} \right)}}{\rm{ + }}\frac{{\omega (1 - {\mu _m}){\lambda _m}}}{F}\left. { + \frac{{\nu(1 - {\mu _m}){\lambda _m}}}{{{D_{m,n}}\left( {\bm{w}_{n,m},{\bm{d}_{n,m}}} \right)}}} \right\},
   \end{aligned}
  \end{array}\label{7}
  \end{equation}
where $\mu_\emph{m}{\lambda _m}$ is the fraction of the task that user $m$ processes
locally (called local computing) with $\mu_\emph{m} \in \left[0, 1\right]$ being the task division parameter. $\frac{{\omega _m}{\mu _m}{\lambda _m}}{f_m}$ represents the computational time of user \emph{m}, $\frac{\omega (1 - {\mu _m}){\lambda _m}}{F}$ represents the time consumption for computing the offloaded task in the MEC server, $\frac{(1-{\mu _m}){\lambda _m}}{U_{n,m}\left( {\bm{v}_{n,m}},{\bm{u}_{n,m}}\right)}$ and $\frac{(1-{\mu _m}){\lambda _m}{\nu}}{D_{n,m}\left( {\bm{w}_{n,m}},{\bm{d}_{n,m}} \right)}$ represent the time for the computational task transmission over uplink and downlink, respectively. In our model, each BS cannot simultaneously communicate with the users and compute the tasks that are offloaded from the users. This is because each BS must first communicate with the users to receive each user's offloaded task and then compute these tasks. Since the collaborative task can be processed by the MEC server and the user simultaneously, $t^3_m$ depends on the maximum time between the local computing time ${\frac{{\omega _m}{\mu _m}{\lambda _m}}{f_m}}$ and the edge computing time $\frac{(1-{\mu _m}){\lambda _m}}{U_{n,m}\left( {\bm{v}_{n,m}},{\bm{u}_{n,m}} \right)}+\frac{\omega (1 - {\mu _m}){\lambda _m}}{F}+\frac{(1-{\mu _m}){\lambda _m}{\nu}}{D_{n,m}\left( {\bm{w}_{n,m}},{\bm{d}_{n,m}} \right)}$, as shown in (7).
  \end{itemize}

\subsection{Problem Formulation}
Next, we formulate the optimization problem that aims to minimize the maximal computational and transmission delay among all users. The minimization problem involves determining uplink subcarrier allocation indicator ${\bm{u}_{n,m}}$, downlink subcarrier allocation indicator ${\bm{d}_{n,m}}$, the uplink transmit power ${\bm{v}_{n,m}}$, the downlink transmit power ${\bm{w}_{n,m}}$, and the task allocation indicator $\mu_\emph{m}$ of each user $m$. The optimization problem can be formulated as follows:
\begin{align}\tag{8}
&\mathop{\rm{min}}\limits_{\mathop {\bm U_{n},\bm V_{n},}\limits_{ \bm D_{n},\bm W_{n},{\bm \mu}} }\!\!\!\! \left({\mathop {\rm{max}}\limits_{m \in {\cal M}} t_m^\phi \left( \bm u_{n,m},\bm v_{n,m},\bm d_{n,m},\bm w_{n,m},{\mu _m}\right)} \right)\!,\\
\tag{8a}
&{~\rm s.t.~}~{\phi} \in \left\{ {1,2,3} \right\},\\
\tag{8b}
&~~~~~~{ u_{n,m}^i, d_{n,m}^j} \in \left\{ {0,1} \right\},\forall n \in \mathcal{N}, \forall m \in \mathcal{M}, \forall i \in \mathcal{I}, \forall j \in \mathcal{J}, \\
\tag{8c}
&~~~~~~\sum\limits_{m \in \mathcal{M}} u_{n,m}^i  \le 1,\forall n \in \mathcal{N},\forall i \in \mathcal{I},\,\\
\tag{8d}
&~~~~~~\sum\limits_{m \in \mathcal{M}} {d_{n,m}^j}  \le 1,\forall n \in \mathcal{N},\forall j \in \mathcal{J},\,\\
\tag{8e}
&~~~~~~\sum\limits_{n \in \mathcal{N}} {u_{n,m}^i}  \le 1,\forall m \in \mathcal{M},\forall {i} \in \mathcal{I},\,\\
\tag{8f}
&~~~~~~\sum\limits_{n \in \mathcal{N}} {d_{n,m}^j}  \le 1,\forall m \in \mathcal{M},\forall {j} \in \mathcal{J},\\
\tag{8g}
&~~~~~~\sum\limits_{i \in \mathcal{I}} {\sum\limits_{m \in \mathcal{M}} {v_{n,m}^i}  \le P_{\rm U},\forall n \in \mathcal{N}} ,\\
\tag{8h}
&~~~~~~\sum\limits_{j \in \mathcal{J}} {\sum\limits_{n \in \mathcal{N}} w_{n,m}^j  \le P_{\rm B},\forall n \in \mathcal{N}},\\
\tag{8i}
&~~~~~~0 \le \mu_{m}  \le 1\, ,
\end{align}
where $\bm U_{n}\!=\![\bm u_{n,1},\ldots,\bm u_{n,M}]$, $\bm V_{n}\!=\![\bm v_{n,1},\ldots,\bm v_{n,M}]$, $\bm D_{n}=[\bm d_{n,1},\!\ldots\!,\bm d_{n,M}]$, $\bm W_{n}\!\!=\!\![\bm w_{n,1},\!\ldots\!,\!\bm w_{n,M}]$, and $\bm \mu\!=\![\mu_1,\!\ldots\!,\mu_{M}]$. (8a) implies that each user can request one of three types of computational tasks. (8b) indicates the uplink and downlink subcarrier allocation between user $m$ and BS $n$. (8c) and (8d) guarantee that each uplink or downlink subcarrier can be allocated to at most one user. (8e) and (8f) ensure that each user can connect to at most one BS for data transmission. (8g) and (8h) are the constraints on the maximum transmit power of each BS $n$ and each user $m$, respectively. (8i) indicates that the collaborative tasks can be cooperatively processed by both BSs and users. Problem (8) is a mixed integer nonlinear programming problem with discrete variables $u_{n,m}^i$ and $d_{n,m}^j$ and continuous variables $v_{n,m}^i$, $w_{n,m}^j$, and $\mu_\emph{m}$. Hence, it is difficult to solve problem (8) by traditional algorithms such as dual method directly \cite{YFNH}. Moveover, as the data size of each computational task requested by each user varies, the BSs must rerun their optimization algorithms to cope with this change thus resulting in additional overhead and delay for computational task processing \cite{MUWC}. In consequence, we develop a novel RL approach that can find a relationship between the users' computational task and resource allocation policy so as to directly generate the resource allocation policy without the time consumption for finding the optimal resource allocation strategy.

\section{Reinforcement Learning for Optimization of Resource Allocation}
Next, we introduce a novel RL approach to solve the optimization problem in (8). First, the components of the proposed learning algorithm is introduced. Then, we explain the use of the learning algorithm to solve (8). Finally, the convergence and implementation of the proposed algorithm is analyzed.

\subsection{Components of Multi-stack RL Method}
A multi-stack RL algorithm consists of three components: a) state, b) action, and c) reward. In particular, $\mathcal{X}$ is the discrete space of environment states, $\mathcal{A}_n^k$ is the discrete sets of available actions for BS \emph{n} at step \emph{k}, and $R$ is the reward function of BS \emph{n}. The components of the multi-stack RL algorithm are specified as follows:

\begin{itemize}
\item \textbf{State}: The environment state $\bm x\in \mathcal{X}$ consists of three components, $\bm x=(t_{\rm max},m_{\rm max},m^*)$, where $t_{\rm max}={\mathop {{\rm{max}}}\limits_{m \in {\cal M}}t_m^\phi\left( \bm u_{n,m},\bm v_{n,m},\bm d_{n,m},\bm w_{n,m},{\mu _m} \right)}$~ represents the maximal computational and transmission delay among all users, $m_{\rm max}=\mathop{\rm{argmax} }\limits_{ m \in {\cal M}} t_m^\phi  \left(\bm u_{n,m},\!\bm v_{n,m},\!\bm d_{n,m},\!\bm w_{n,m},\!{\mu _m} \right)$ represents the user whose time consumption is maximal among all users, and $m^*$ represents the notion of the user that requests computational and transmission resource at current step. Note that, $t_{\rm max}$ is determined by the finite and discrete actions $\bm u_{n,m}$, $\bm v_{n,m}$, $\bm d_{n,m}$, $\bm w_{n,m}$ and ${\mu _m}$. Since $m_{\rm max}\in \{1,\ldots,M\}$ and $m^* \!\in\! \{1,\ldots,M\}$, the defined environment states are finite and discrete.

\item \textbf{Action}: Since each BS jointly optimizes task, subcarrier, and transmit
power allocation scheme, the action $\bm a_{n}^k=[\bm u_{n},\bm v_{n},\bm d_{n},\bm w_{n}]$, where $\bm u_{n}\!=\![\bm u_{1,n},\ldots,\bm u_{M,n}]$, $\bm v_{n}\!=\![\bm v_{1,n},\ldots,\bm v_{M,n}]$, $\bm d_{n}=[\bm d_{1,n},\ldots,\bm d_{M,n}]$, and $\bm w_{n}=[\bm w_{1,n},\ldots,\bm w_{M,n}]$. The uplink transmit power and downlink transmit power are separately divided into $N_a$ levels. Hence, we assume that $v_{n,m}^i \in \{0,\frac{P_{\rm U}}{N_a},\frac{2P_{\rm U}}{N_a},\ldots,P_{\rm U}\}$ and $w_{n,m}^i \in \{0,\frac{P_{\rm B}}{N_a},\frac{2P_{\rm B}}{N_a},\ldots,P_{\rm B}\}$. To find the optimal task allocation $\mu_m$, we present the following result:
\renewcommand{\arraystretch}{1.5}
\begin{table*}[t]
  \centering
  \fontsize{7.5}{11}\selectfont

  \caption{\quad Summarization of the Time Consumption.}

  \label{tab:performance_comparison}

    \begin{tabular}{|c|c|c|c|c|}

    \hline

    \multirow{2}{*}{The type of computational tasks}&

    \multicolumn{4}{c|}{The variation of resource allocation}\cr\cline{2-5}

    &downlink subcarriers&uplink subcarriers&downlink transmit power&uplink transmit power\cr

    \hline

    Edge task&$\Delta t_m^1(\bm{d}_{n,m}\!+\!\Delta \bm{d}_{n,m})$&$\Delta t_m^1(\bm{u}_{n,m}\!+\!\Delta \bm{u}_{n,m})$&$\Delta t_m^1(\bm{w}_{n,m}\!+\!\Delta \bm{w}_{n,m})$&$\Delta t_m^1(\bm{v}_{n,m}\!+\!\Delta \bm{v}_{n,m})$\cr\hline

    Local task&$\Delta t_m^2(\bm{d}_{n,m}\!+\!\Delta \bm{d}_{n,m})$&$\Delta t_m^2(\bm{u}_{n,m}\!+\!\Delta \bm{u}_{n,m})$&$\Delta t_m^2(\bm{w}_{n,m}\!+\!\Delta \bm{w}_{n,m})$&$\Delta t_m^2(\bm{v}_{n,m}\!+\!\Delta \bm{v}_{n,m})$\cr\hline

    Collaborative task&$\Delta t_m^3(\bm{d}_{n,m}\!+\!\Delta \bm{d}_{n,m})$&$\Delta t_m^3(\bm{u}_{n,m}\!+\!\Delta \bm{u}_{n,m})$&$\Delta t_m^3(\bm{w}_{n,m}\!+\!\Delta \bm{w}_{n,m})$&$\Delta t_m^3(\bm{v}_{n,m}\!+\!\Delta \bm{v}_{n,m})$\cr\hline

    \end{tabular}
    \vspace{-0.8cm}
\end{table*}

\begin{theorem}\label{thm1}{\rm For the collaborative task, the optimal task allocation is given by:}
\end{theorem}
\vspace{-0.1cm}
\begin{equation}\tag{9}
\begin{aligned}
&{\mu _m} = \frac{\omega f_m+Y}{\omega f_m+Y+\omega_m F},\
\end{aligned}\label{7}
\end{equation}
where ${Y}{\rm{ = }}\frac{{f_m}F}{{U_{n,m}}({\bm{v}_{n,m}},{\bm{u}_{n,m}})}+\frac{{f_m}\nu F}{{D_{n,m}}({\bm{w}_{n,m}},{\bm{d}_{n,m}})}.$
\vspace{0.2cm}
\begin{IEEEproof}See Appendix A.
\end{IEEEproof}\vspace{0.2cm}
\quad Theorem 1 shows that the task allocation depends on the transmit power and the subcarrier allocation. In particular, as the transmit power and the number of the subcarriers over uplink and downlink allocated to each user increases, the part of a task computed by the MEC server increases. In consequence, the computational time decreases.

\quad Substituting (9) into (7), we have:
\begin{equation}\tag{10}
\begin{aligned}
{t}_m^3({\bm{u}_{n,m}},{\bm{v}_{n,m}},{\bm{d}_{n,m}},{\bm{w}_{n,m}},{\mu _m})\!=\!\frac{{\omega _m}{\lambda _m}({\omega f_m\!\!+\!\!Y})}{f_m(\omega f_m\!\!+\!\!Y\!\!+\!\omega_m F)},
\end{aligned}
\end{equation}
where ${Y}{\rm{ = }}\frac{{f_m}F}{{U_{n,m}}({\bm{v}_{n,m}},{\bm{u}_{n,m}})}+\frac{{f_m}\nu F}{{D_{n,m}}({\bm{w}_{n,m}},{\bm{d}_{n,m}})}.$

\vspace{0.3cm}
In Theorem 1, we build the relationship between the task allocation and the transmit power and the subcarrier allocation. Next, we analyze the gain that stems from the change of the transmit power and the number of the subcarriers over uplink and downlink allocated to user \emph{m}. To present the reduction of the delay due to the change of the number of the subcarriers and transmit power allocated to user, we first summarize the time consumption notations, as shown in Table I. In Table I, $\Delta t_m^1$ represents the variation of time consumption for processing edge task when the resource allocation scheme changes. In particular, $\Delta t_m^1(\bm{d}_{n,m}\!+\!\Delta \bm{d}_{n,m})$, $\Delta t_m^1(\bm{u}_{n,m}\!+\!\Delta \bm{u}_{n,m})$, $\Delta t_m^1(\bm{w}_{n,m}\!+\!\Delta \bm{w}_{n,m})$, and $\Delta t_m^1(\bm{v}_{n,m}\!+\!\Delta \bm{v}_{n,m})$, respectively, represents the variation of time consumption for processing edge task due to the change of the number of downlink subcarriers, the number of uplink subcarriers, downlink transmit power, and uplink transmit power. Similarly, $\Delta t_m^2$ and $\Delta t_m^3$ represent the variation of time consumption for processing local task and collaborative task when the resource allocation scheme changes, respectively. Given time consumption notions, we present the relationship between the time consumption and the change of the number of subcarriers and transmit power allocated to each user.

\begin{theorem}\label{thm1}{\rm The reduction of the delay due to the change of the number of the subcarriers and transmit power allocated to user \emph{m} is:}
\vspace{-0.2cm}
\end{theorem}
\begin{itemize}
\item The gain due to the change of the number of downlink subcarriers allocated to user \emph{m} that requests an edge task, $\Delta {t}_m^1({\bm{w}_{n,m},\bm{d}_{n,m}}\!+\!\Delta\bm{d}_{n,m})$, is:
\begin{equation}\tag{11}
\begin{aligned}
\!\!\!\!\!\!\Delta &{t}_m^1({\bm{w}_{n,m},\bm{d}_{n,m}}\!+\!\Delta\bm{d}_{n,m})
=\!\!&\left\{ \begin{array}{l}
\!\!\!\!\frac{{\nu{\lambda _m}}}{{{D_{n,m}}\left( {\bm{w}_{n,m},\bm{d}_{n,m}} \right)}}, \left\|\Delta \bm{d}_{n,m}\right\| \!\gg\! \left\|\bm{d}_{n,m}\right\|,\\
\!\!\!\!\frac{{\nu{\lambda _m}\Delta \bm{d}_{n,m}}}{{\bm{d}_{n,m}^2{D_{n,m}}\left( {\bm{w}_{n,m},\bm{d}_{n,m}} \right)}}, \left\|\Delta \bm{d}_{n,m}\right\| \!\ll\! \left\|\bm{d}_{n,m}\right\|,\\
\!\!\!\!\frac{{\nu{\lambda _m}\Delta \bm{d}_{n,m}}}{{\bm{d}_{n,m}(\bm{d}_{n,m} + \Delta \bm{d}_{n,m}){D_{n,m}}\left( {\bm{w}_{n,m},\bm{d}_{n,m}} \right)}},\rm{else},
\end{array} \right.\
\end{aligned}
\end{equation}where $\Delta \bm{d}_{n,m}\!\!=\!\![\Delta d_{n,m}^{1},\!\Delta d_{n,m}^{2},\!\ldots\!,\!\Delta d_{n,m}^{J}]$ represents the variation of downlink subcarriers allocation. $\Delta d_{n,m}^{j}\!=\!1$ indicates that BS $n$ allocates downlink subcarrier $j$ to user $m$, otherwise, we have $\Delta d_{n,m}^{j}\!=\!0$. $\left\| {\Delta\bm d_{n,m}}\right\|$ is~the module of ${\Delta\bm d_{n,m}}$, which indicates the number of downlink subcarriers that will be allocated to user $m$. Similarly, $\left\|{\bm d_{n,m}}\right\|$ represents the number of downlink subcarriers that are already allocated to user $m$.
\end{itemize}
\vspace{-0.2cm}
\begin{itemize}
\item The gain that stems from the change of the number of uplink subcarriers allocated to user \emph{m} that requests a local task, $\Delta {t}_m^2({\bm{v}_{n,m},\bm{u}_{n,m}}\!+\!\Delta\bm{u}_{n,m})$, is:
\begin{equation}\tag{12}
\begin{aligned}
\!\!\Delta & {t}_m^2({\bm{v}_{n,m},\bm{u}_{n,m}}\!+\!\Delta\bm{u}_{n,m})
=\!\!&\left\{ \begin{array}{l}
\!\!\!\!\frac{{\nu{\lambda _m}}}{{{U_{n,m}}\left( {\bm{v}_{n,m},\bm{u}_{n,m}} \right)}},\left\|\Delta \bm{u}_{n,m}\right\| \!\gg\! \left\|\bm{u}_{n,m}\right\|,\\
\!\!\!\!\frac{{\nu{\lambda _m}\Delta \bm{u}_{n,m}}}{{\bm{u}_{n,m}^2{U_{n,m}}\left( {\bm{v}_{n,m},\bm{u}_{n,m}} \right)}},\left\|\Delta \bm{u}_{n,m}\right\| \!\ll\! \left\|\bm{u}_{n,m}\right\|,\\
\!\!\!\!\frac{{\nu{\lambda _m}\Delta \bm{u}_{n,m}}}{{\bm{u}_{n,m}(\bm{u}_{n,m} + \Delta \bm{u}_{n,m}){U_{n,m}}\left( {\bm{v}_{n,m},\bm{u}_{n,m}} \right)}},{\rm else},
\end{array} \right.\
\end{aligned}
\end{equation}
where $\Delta \bm{u}_{n,m}\!\!=\!\![\Delta u_{n,m}^{1}, \Delta u_{n,m}^{2}, \ldots, \Delta u_{n,m}^{I}]$ represents the variation of uplink subcarriers allocation. Similarly, $\Delta u_{n,m}^{i}=1$ indicates that BS $n$ allocates uplink subcarrier $i$ to user $m$ and $\Delta u_{n,m}^{i}=0$, otherwise. $\left\| {\Delta\bm u_{n,m}}\right\|$ indicates the number of uplink subcarriers that will be allocated to user $m$. $\left\|{\bm u_{n,m}}\right\|$ is the number of uplink subcarriers that are already allocated to user $m$.
\end{itemize}
\vspace{-0.1cm}
\begin{itemize}
\item The gain that stems from the change of the number of downlink subcarriers allocated to user \emph{m} that requests a collaborative task, $\Delta {t}_m^3({\bm{w}_{n,m},\bm{d}_{n,m}}\!+\!\Delta\bm{d}_{n,m})$, is:
\begin{equation}\tag{13}
\begin{aligned}
&\Delta {t}_m^3({\bm{w}_{n,m},\bm{d}_{n,m}}\!+\!\Delta\bm{d}_{n,m})= &\!\!\left\{ \begin{array}{l}
\!\!\!\!\frac{\lambda _m\nu(\omega_m F)^2}{{(\omega_m F\!+\! Y)(\omega_m F\!+\!\Delta Y_d)}{D_{\!n\!,\!m}}\left( {\bm w_{n\!,\!m},\bm d_{n\!,\!m}} \right)} ,\!\left\| \Delta \bm{d}_{n,m} \right\| \!\!\gg\!\! \left\|\bm{d}_{n,m}\right\| \!,\\
\!\!\!\!\!\frac{{\lambda _m}\nu(\omega_m F)^2\Delta \bm d_{n\!,\!m}}{\bm d_{n\!,\!m}^2\!(\omega_m F\!+\! Y)(\omega_m F\!+\!\Delta Y_d){D_{\!n\!,\!m}}\left( {\bm w_{n\!,\!m},\bm d_{n\!,\!m}}\! \right)},\!\left\| \Delta \bm{d}_{n,m} \right\| \!\!\ll\!\! \left\| \bm{d}_{n,m} \right\| \!,\\\vspace{0.1cm}
\!\!\!\!\!\frac{{\lambda _m}\nu(\omega_m F)^2\Delta \bm d_{n\!,\!m}}{\bm d_{n\!,\!m}(\bm d_{n\!,\!m} \!+\! \Delta \bm d_{n\!,\!m})(\omega_m F\!+\! Y)(\omega_m F\!+\!\Delta Y_d){D_{\!n\!,\!m}}\!\left(\! {\bm w_{n,m},\bm d_{n\!,\!m}}\! \right)} ,\!{\rm else},
\end{array} \right.\
\end{aligned}
\end{equation}
where $Y{\rm{ = }}\frac{{f_m}F}{{U_{n,m}}(\bm{v}_{n,m},\bm{u}_{n,m})}+\frac{{f_m}\nu F}{{D_{n,m}}(\bm{w}_{n,m},\bm{d}_{n,m})}$ and $\Delta Y_d=\frac{{f_m}F}{{U_{n,m}}(\bm{v}_{n,m},\bm{u}_{n,m})}+\frac{{f_m}\nu F}{{D_{n,m}}(\bm{w}_{n,m},\bm{d}_{n,m}+\Delta\bm{d}_{n,m})}.$
\end{itemize}
\vspace{-0.1cm}
\begin{itemize}
\item The gain that stems from the change of the number of uplink subcarriers allocated to user \emph{m} that requests a collaborative task, $\Delta {t}_m^3({\bm{v}_{n,m},\bm{u}_{n,m}}\!+\!\Delta\bm{u}_{n,m})$, is:
\begin{equation}\tag{14}
\begin{aligned}
&\Delta {t}_m^3({\bm{v}_{n,m},\bm{u}_{n,m}}\!+\!\Delta\bm{u}_{n,m})\!= &\!\!\! \left\{ \begin{array}{l}
\!\!\!\! \frac{\lambda_m(\omega_m F)^2}{(\omega_m F\!+\!Y)(\omega_m F\!+\!\Delta Y_u){D_{\!n\!,\!m}}\!\left(\! {\bm w_{n,m},\bm d_{n\!,\!m}} \!\right)},\!\left\| \Delta \bm{u}_{n,m} \right\| \!\!\gg\!\! \left\|\bm{u}_{n,m} \right\|\! ,\\
\!\!\!\!\!\frac{{\lambda _m}(\omega_m F)^2\Delta \bm d_{n\!,\!m}}{{\bm d_{n\!,\!m}^2\!(\omega_m F\!+\! Y)(\omega_m F\!+\!\Delta Y_u){D_{\!n\!,\!m}}\left( {\bm w_{n,m},\bm d_{n\!,\!m}} \right)}},\!\left\| \Delta \bm{u}_{n,m} \right\| \!\!\ll\!\! \left\| \bm{u}_{n,m} \right\|\! ,\\
\!\!\!\! \frac{{\lambda _m}(\omega_m F)^2\Delta \bm d_{n\!,\!m}}{{\bm d_{n\!,\!m}(\bm d_{n\!,\!m} \!+\!\Delta \bm d_{n\!,\!m}\!)(\omega_m F\!+\! Y)(\omega_m F\!+\!\Delta Y_u){D_{\!n\!,\!m}}\!\left(\! {\bm w_{n\!,\!m},\bm d_{n\!,\!m}}\! \right)}},{\rm else},
\end{array} \right.\
\end{aligned}
\end{equation}
where $Y{\rm{ = }}\frac{{f_m}F}{{U_{n,m}}(\bm{v}_{n,m},\bm{u}_{n,m})}+\frac{{f_m}\nu F}{{D_{n,m}}(\bm{w}_{n,m},\bm{d}_{n,m})}$ and $\Delta Y_u{\rm{ = }} \frac{{f_m}F}{{U_{n,m}}(\bm{v}_{n,m},\bm{u}_{n,m}+\Delta\bm{u}_{n,m})}+\frac{{f_m}\nu F}{{D_{n,m}}(\bm{w}_{n,m},\bm{d}_{n,m})}.$
\end{itemize}
\begin{itemize}
\item The gain that stems from the change of the downlink transmit power of \emph{m} that requests an edge task, $\Delta {t}_m^1({\bm{w}_{n,m}\!+\!\Delta\bm{w}_{n,m},\!\bm{d}_{n,m}})$, is:
\begin{equation}\tag{15}
\begin{aligned}
&\Delta {t}_m^1({\bm{w}_{n,m}\!+\!\Delta\bm{w}_{n,m},\!\bm{d}_{n,m}})\!=\!
& \frac{\nu{\lambda _m}({{D_{n,m}}\!\left( {\bm{w}_{n,m}\!\!+\!\!\Delta \bm{w}_{n,m},\!\bm{d}_{n,m}} \right)}\!\!-\!\!{{D_{n,m}}\!\left( {\bm{w}_{n,m},\!\bm{d}_{n,m}} \right))}}{{{D_{n,m}}\!\left( {\bm{w}_{n,m},\!\bm{d}_{n,m}} \right)}{{D_{n,m}}\!\left( {\bm{w}_{n,m}\!\!+\!\!\Delta \bm{w}_{n,m},\!\bm{d}_{n,m}} \right)}}.\\
\end{aligned}
\end{equation}
\end{itemize}
\vspace{-0.1cm}
\begin{itemize}
\item The gain that stems from the change of the uplink transmit power of \emph{m} that requests a local task, $\Delta {t}_m^2({\bm{w}_{n,m}\!+\!\Delta\bm{w}_{n,m},\!\bm{u}_{n,m}})$, is:
\begin{align}
&\Delta {t}_m^2({\bm{w}_{n,m}\!+\!\Delta\bm{w}_{n,m},\!\bm{u}_{n,m}})\!=\!
&\frac{\nu{\lambda _m}({{U_{n,m}}\!\left( {\bm{v}_{n,m}\!\!+\!\!\Delta \bm{v}_{n,m},\!\bm{u}_{n,m}} \right)}\!-\!{{U_{n,m}}\!\left( {\bm{v}_{n,m},\!\bm{u}_{n,m}} \right))}}{{{U_{n,m}}\!\left( {\bm{v}_{n,m},\!\bm{u}_{n,m}} \right)}{{U_{n,m}}\!\left( {\bm{v}_{n,m}\!\!+\!\!\Delta \bm{v}_{n,m},\!\bm{u}_{n,m}} \right)}}. \tag{16}
\end{align}

\end{itemize}
\vspace{-0.3cm}
\begin{itemize}
\item The gain that stems from the change of the downlink transmit power of user \emph{m} that requests a collaborative task, $\Delta {t}_m^3({\bm{w}_{n,m}\!+\!\Delta\bm{w}_{n,m},\!\bm{d}_{n,m}})$, is:
\begin{equation}\tag{17}
\begin{aligned}
&\Delta {t}_m^3({\bm{w}_{n,m}\!+\!\Delta\bm{w}_{n,m},\!\bm{d}_{n,m}})\\
&\!=\!\frac{{\lambda _m}\nu(\omega_m F)^2}{(\omega_m F\!\!+\!\!Y)(\omega_m F\!\!+\!\!\Delta Y_w)}
\times\! \frac{{{D_{n,m}}\!\left( {\bm{w}_{n,m}\!\!+\!\!\Delta \bm{w}_{n,m},\!\bm{d}_{n,m}} \right)}\!-\!{{D_{n,m}}\!\left( {\bm{w}_{n,m},\!\bm{d}_{n,m}} \right)}}{{{D_{n,m}}\!\left( {\bm{w}_{n,m},\!\bm{d}_{n,m}} \right)}{{D_{n,m}}\!\left( {\bm{w}_{n,m}\!\!+\!\!\Delta \bm{w}_{n,m},\!\bm{d}_{n,m}} \right)}}\!,
\end{aligned}
\end{equation}
where $Y{\rm{ = }}\frac{{f_m}F}{{U_{n,m}}(\bm{v}_{n,m},\bm{u}_{n,m})}+\frac{{f_m}\nu F}{{D_{n,m}}(\bm{w}_{n,m},\bm{d}_{n,m})}$ and $\Delta Y_w{\rm{ = }}\frac{{f_m}F}{{U_{n,m}}(\bm{v}_{n,m},\bm{u}_{n,m})}+\frac{{f_m}\nu F}{{D_{n,m}}(\bm{w}_{n,m}+\Delta\bm{w}_{n,m},\bm{d}_{n,m})}.$
\end{itemize}
\begin{itemize}
\item The gain that stems from the change of the uplink transmit power of user \emph{m} that requests a collaborative task, $\Delta {t}_m^{3}({\bm{v}_{n,m}\!\!+\!\!\Delta\bm{v}_{n,m},\!\bm{u}_{n,m}})$, is:
    \vspace{-0.2cm}
\begin{equation}\tag{18}
\begin{aligned}
&\Delta {t}_m^3({\bm{v}_{n,m}\!\!+\!\!\Delta\bm{v}_{n,m},\!\bm{u}_{n,m}})\\
&\!=\!\frac{{\lambda _m}(\omega_m F)^2}{(\omega_m F\!\!+\!\!Y)(\omega_m F\!\!+\!\!\Delta Y_u)}\times\!\! \frac{{{U_{n,m}}\!\left( {\bm{v}_{n,m}\!\!+\!\!\Delta \bm{v}_{n,m},\!\bm{u}_{n,m}} \right)}\!-\!{{U_{n,m}}\!\left( {\bm{v}_{n,m},\!\bm{u}_{n,m}} \right)}}{{{U_{n,m}}\!\left( {\bm{v}_{n,m},\!\bm{u}_{n,m}} \right)}{{U_{n,m}}\!\left( {\bm{v}_{n,m}\!\!+\!\!\Delta \bm{v}_{n,m},\!\bm{u}_{n,m}} \right)}}\!,
\end{aligned}
\end{equation}
where ${Y}{\rm{ = }}\frac{{f_m}F}{{U_{n,m}}(\bm{v}_{n,m},\bm{u}_{n,m})}+\frac{{f_m}\nu F}{{D_{n,m}}(\bm{w}_{n,m},\bm{d}_{n,m})}$ and $\Delta Y_v{\rm{ = }}\frac{{f_m}F}{{U_{n,m}}(\bm{v}_{n,m}+\Delta\bm{v}_{n,m},\bm{u}_{n,m})}+\frac{{f_m}\nu F}{{D_{n,m}}(\bm{w}_{n,m},\bm{d}_{n,m})}.$
\end{itemize}
\vspace{0.1cm}
\begin{IEEEproof}See Appendix B.
\end{IEEEproof}
\vspace{0.1cm}

\quad From Theorem 2, we can see that the number of subcarriers and transmit power allocated to each user \emph{m}, will directly affect the delay of user \emph{m}. Therefore, to minimize the maximal transmission and computational delay among users, we can increase the number of subcarriers as well as the transmit power allocated to each user according to the type of the task that each user requests. Although increasing the number of subcarriers as well as the transmit power allocated to each user can decrease the delay of each user, the gain that stems from  increasing the same number of subcarriers or transmit power allocated to the user who requests various types of computational tasks is different. To capture the maximum gain that stems from the change of the same number of subcarriers and the transmit power as a given user has various types of computational tasks, we state the following result:
\vspace{-0.1cm}
\begin{corollary}\label{1.1}{\rm The relationship among the gains that stem from the change of the same number of subcarriers or transmit power for a user that has different computational tasks are:
\begin{itemize}
\item The relationship among the gains that stem from the change of the number of downlink subcarriers allocated to user \emph{m} is: $\Delta {t}_m^{2}(\bm{d}_{n,m}+\Delta \bm{d}_{n,m})<\Delta {t}_m^{3}(\bm{d}_{n,m}+\Delta \bm{d}_{n,m})<\Delta {t}_m^{1}(\bm{d}_{n,m}+\Delta \bm{d}_{n,m})$.
\end{itemize}
\begin{itemize}
\item The relationship among the gains that stem from the change of the number of uplink subcarriers allocated to user \emph{m} is: $\Delta {t}_m^{1}(\bm{u}_{n,m}+\Delta \bm{u}_{n,m})<\Delta {t}_m^{3}(\bm{u}_{n,m}+\Delta \bm{u}_{n,m})<\Delta {t}_m^{2}(\bm{u}_{n,m}+\Delta \bm{u}_{n,m})$.
\end{itemize}
\begin{itemize}
\item The relationship among the gains that stem from the change of the downlink transmit power allocated to user \emph{m} is: $\Delta t_m^{2}(\bm{w}_{n,m}+\Delta \bm{w}_{n,m})<\Delta t_m^{3}(\bm{w}_{n,m}+\Delta \bm{w}_{n,m})<\Delta t_m^{1}(\bm{w}_{n,m}+\Delta \bm{w}_{n,m})$.
\end{itemize}
\begin{itemize}
\item The relationship among the gains that stem from the change of the uplink transmit power allocated to user \emph{m} is: $\Delta t_m^{1}(\bm{v}_{n,m}+\Delta \bm{v}_{n,m})<\Delta t_m^{3}(\bm{v}_{n,m}+\Delta \bm{v}_{n,m})<\Delta t_m^{2}(\bm{v}_{n,m}+\Delta \bm{v}_{n,m})$.
\end{itemize}
\begin{IEEEproof}See Appendix C.
\end{IEEEproof}
\vspace{0.1cm}
\quad From Corollary 1, we can see that, the gain that stems from increasing the number of subcarriers and the transmit power of a user who has a collaborative task is less than that for a user that requests an edge task or a local task. This is because as the number of subcarriers or transmit power for uplink (downlink) increases, the data rate for uplink (downlink) increases, thus decreasing the uplink (downlink) transmission delay. Meanwhile, due to the increase of the uplink (downlink) transmission rate, the user will send more data to the MEC server that can use its high performance CPUs to process the data. Thus, the downlink (uplink) transmission delay increases. In particular, the increase of the downlink (uplink) transmission delay is lager than the decrease of the computational delay. Based on Theorem 2 and Corollary 1, to minimize the maximal computation and transmission delay among all users, BS \emph{n} prefers to allocate more downlink subcarriers and downlink transmit power to a user that requests an edge task and allocate more uplink subcarriers and uplink transmit power to a user that requests a local task}.
\end{corollary}

\item \textbf{Reward}: Given the current environment state $\bm x$ and the selected action $\bm a_n^k$, the reward function of each BS $n$ is given by:
\begin{equation}\tag{19}
\begin{split}
&R(\bm x,\!\bm a_n^k)\!=\!\frac{\mathop {{\rm{max}}}\limits_{m \in {\cal M}} {{\frac{\lambda _m}{f_m}}\!-\!\mathop {{\rm{max}}}\limits_{ n \in {\cal N}} {t^\phi_m(\bm a_n^k)}}}{\mathop {{\rm{max}}}\limits_{m \in {\cal M}} {\frac{\lambda _m}{f_m}}},
\end{split}
\vspace{-0.2cm}
\end{equation}
where $R(\bm x,\emph{\textbf{a}$_n^k$})\in\left(0, 1\right)$ with $\mathop {{\rm{max}}}\limits_{m \in {\cal M}}\!{{\frac{\lambda _m}{f_m}}}$ being~the maximal time consumption of all users to process its own task locally and $\mathop {{\rm{max}}}\limits_{n \in {\cal N}} {t^\phi_m(\bm a_n^k)}$ being the maximal transmission and computational time of all users. To calculate $\mathop {{\rm{max}}}\limits_{n \in {\cal N}} {t^\phi_m(\bm a_n^k)}$, each BS $n$ must exchange its maximal delay among its associated users with other BSs so as to adjust the resource allocation scheme to minimize the maximal computational and transmission delay among all users.
\end{itemize}
\subsection{Multi-stack RL for Optimization of Resource Allocation}
Given the components of the proposed learning algorithm (the flowchart is shown in Algorithm 1), next, we present the use of the proposed learning algorithm to solve problem (8). In particular, each BS \emph{n} first selects an action \textbf{\emph{a$_n^k$}} from $\mathcal{A}_n^k$ at each step $k$. After the selected action \textbf{\emph{a$_n^k$}} is performed by BS \emph{n}, the environment state $\bm x$ changes and BS $n$ records the obtained reward $R(\bm x,\emph{\textbf{a}$_n^k$})$ in its Q-table \emph{Q}($\bm x$, \textbf{\emph{a$^k_n$}}). To ensure that any action can be chosen with a non-zero probability, an $\epsilon$-greedy exploration \cite{YFYR} is adopted. This mechanism is responsible for action selection during the learning process and balance the tradeoff between exploration and exploitation.
Here, exploration refers to the case in which each BS explores actions to find a better strategy. Exploitation refers to the case in which each BS will adopt the action with the maximum reward. Therefore, the probability for BS \emph{n} selecting action \textbf{\emph{a}}$^{k}_n$ can be given by:
\vspace{0.1cm}
\begin{equation}\tag{20}
\begin{aligned}
{\pi _{n,\bm{a^k_n}}} = \left\{ \begin{array}{l}
1 - \varepsilon  + \frac{\varepsilon }{{\left\| {\textbf{\emph{a}}^k_n} \right\|}},{\kern 9pt}\mathop {\arg \max }\limits_{\bm{a}^{k}_{n} \in {\mathcal{A}^k_n}} Q(\bm x,\bm{a}^{k}_{n}),\vspace{0.4cm}\\
\frac{\varepsilon }{{\left\| {\textbf{\emph{a}}^k_n} \right\|}},{\kern 42pt}{\rm otherwise},
\end{array} \right.\
\end{aligned}
\end{equation}
where $\epsilon$ is the probability of exploration.

To avoid repeating the historical resource allocation schemes, the multiple stacks are used to record the information of current resource allocation scheme and users' states, which defined as $\bm v_{0}=\left[v_{0}^1, v_{0}^2, \ldots, v_{0}^B\right]$, $\bm v_{1}=\left[v_{1}^1, v_{1}^2, \ldots, v_{1}^B\right]$, $\ldots$, $\bm v_{G-1}=\left[v_{G-1}^1, v_{G-1}^2, \ldots , v_{G-1}^B\right]$ with \emph{G} being the number of stacks and \emph{B} being the length of each stack. Since the selected action \textbf{\emph{a$^k_n$}} and the current state $\bm x$ will be recorded in element v$_{\rm G}^l$ of the corresponding stack \textbf{v}$_{\rm G}$, the proposed algorithm enables each BS $n$ to learn the information in the stacks, thus increasing the probability of exploration in the first $G \times B$ steps. Then, the selected action and the current state are compared with the historical information that is recorded in the corresponding stack. The comparison process is given by:
\begin{equation}\tag{21}
 \begin{aligned}
 q =&\mathbbm{1}_{\left\{ {k {\kern 2pt}{\rm mod {\kern 2pt}\emph{G}=0}{\kern 2pt} \& {\kern 2pt}{\emph{R}(\bm x,{\kern 2pt}\textbf{\emph{a$^k_n$}})} \ne {v}_{0}^b, b=1,...,B} \right\} }\\
  & \vee \mathbbm{1}_{\left\{ {k {\kern 2pt}{\rm mod {\kern 2pt}\emph{G}=1}{\kern 2pt} \& {\kern 2pt}{\emph{R}(\bm x,{\kern 2pt}\textbf{\emph{a$^k_n$}})} \ne {v}_{1}^b, b=1,...,B} \right\} }\\
  & \vee ...\\
  & \vee \mathbbm{1}_{\left\{ {k {\kern 1pt} {\kern 1pt}{\rm mod} {\kern 1pt}{\kern 1pt}\emph{G}=\emph{G}-1{\kern 2pt} \& {\kern 2pt}{\emph{R}(\bm x,{\kern 1pt}\textbf{\emph{a$^k_n$}})} \ne {v}_{\emph{G}-1}^b},b=1,2,...,B \right\}},\
  \end{aligned}
 \end{equation}
where $\mathbbm{1}_{\{x\}}=1$ as $x$ is true, otherwise, we have $\mathbbm{1}_{\{x\}}=0$. $(k{\kern 3pt} {\rm{mod }}{\kern 3pt} \emph{G}=i)$ indicates that the information obtained by BS $n$ at step \emph{k} must be compared with the records in stack \emph{i}. In addition, $\mathbbm{1}_{\left\{ {k {\kern 2pt}{\rm mod} {\kern 2pt}\emph{G}=i{\kern 2pt} \& {\kern 2pt}{\emph{R}(\bm x,{\kern 1pt}\textbf{\emph{a$^k_n$}})} \ne {v}_{i}^l},b=1,...,B\right\}}=1$ indicates that the information obtained at step $k$ is not recorded in stack $i$. Hence, $q \in \left\{0,1 \right\}$ is the comparison result with $q=1$ indicating that the information at step $k$ is recorded in stack $i$, and $q=0$, otherwise. Next, BS \emph{n} records \emph{R$^{}$}($\bm x$, \textbf{\emph{a$^k_n$}}) in one of its multiple stacks, which is given by:
\begin{equation}\tag{22}
\begin{aligned}
\left\{ \begin{array}{l}
{{v}}_{1}^{1+(k-1)/\emph{G}} {\kern 1pt} {\kern 2pt}=  R(\bm x,\textbf{\emph{a$^k_n$}}),{\kern 5pt} {\rm if}{\kern 3pt} k {\kern 3pt} {\kern 1pt}\rm mod {\kern 1pt}{\kern 1pt}\emph{G}=1{\kern 1pt},\\
{{v}}_{2}^{1+(k-2)/\emph{G}} {\kern 1pt} {\kern 2pt}=  R(\bm x,\textbf{\emph{a$^k_n$}}),{\kern 5pt} {\rm if}{\kern 3pt} k {\kern 3pt} {\kern 1pt}\rm mod {\kern 1pt}{\kern 1pt}\emph{G}=2{\kern 1pt},\\
{\ldots}\\
{{v}}_{G-1}^{1+(k-(G-1))/\emph{G}} {\kern 1pt} {\kern 2pt}=  R(\bm x,\textbf{\emph{a$^k_n$}}),{\kern 5pt} {\rm if}{\kern 3pt} k {\kern 3pt} {\kern 1pt}\rm mod {\kern 1pt}{\kern 1pt}\emph{G}=\emph{G}-1,\\
{{v}}_{\emph{G}}^{k/\emph{G}} =  {R(\bm x,\textbf{\emph{a$^k_n$}})},{\kern 4pt} {\rm if} {\kern 2pt} k {\kern 2pt}\rm mod{\kern 2pt}\emph{G}=0{\kern 1pt},\\
\end{array} \right.\
\end{aligned}
\end{equation}where ${v}_{i}^{1+(k-1)/G}$ indicates that the information obtained at step $k$ should be stored in element ${1+(k-1)/G}$ of stack $i$.

After the information at step $k$ is recorded in the stacks, each BS $n$ will obtain comparison result $q$, reward $R(\bm x, \bm a_n^{k})$, and state $\bm x$ so as to update its Q-table, which can be given by:
\begin{equation}\tag{23}
\begin{aligned}
\setlength{\abovedisplayshortskip}{-0.2cm}
&Q(\bm x,\bm a_n^k)\!
=Q(\bm x,\bm a_n^k)\!+\!\emph{q}\!*\!\alpha (R(\bm x, \bm a_n^k)\!+\!\gamma\mathop {\max }\limits_{\bm a_n^{k'}} Q(\bm x'\!,\textbf{\emph{a$^k_n$}}') \!-\!Q(\bm x,\bm a_n^{k})),\
\end{aligned}
\vspace{-0.1cm}
\end{equation}where $\bm x'$ and $\bm a_n^{k'}$ are the next state and action. $\alpha\!\in\! [0,1]$ is the learning rate and $\gamma\!\in\! [0,1]$ is the discount factor. According to (21)-(23), at each step, each BS must update its Q-table according to the comparison result and record the new resource allocation scheme and users' state in the stacks. Based on the recorded historical information, each BS can avoid repeatedly adopting the same resource allocation scheme thus speeding up convergence. In essence, at each step, BS $n$ chooses an action $\bm a_n^k$ based on $\epsilon$-greedy mechanism and the historical resource allocation schemes so as to allocate its subcarrier and power to its associated users and divide the collaborative task requested by the users. Then, BS $n$ can calculate the maximal computation and transmission delay of its associated users and exchange it with other BSs. Next, each BS can obtain the maximal delay among all users so as to calculate the reward and record the information of its own action and the environment state. Based on the current state $\bm x$, the current reward $R\!\left(\bm x, \bm a_n^k\right)$, and action \textbf{\emph{a$^k_n$}}, BS \emph{n} can update its Q-table.

\begin{algorithm}[t]\small
\caption{Multi-stack RL Method}
\label{table}
\begin{algorithmic} [1] 
\REQUIRE The available action space $\mathcal{A}_n^k$ and the environment state $\mathcal{X}$.\\
\ENSURE The resource allocation policy.\\ 
\STATE Initialize the stacks \emph{\textbf{v}$_{1}$}, \emph{\textbf{v}$_{2}$}, \ldots, \emph{\textbf{v}$_{G}$} and \emph{Q}($\bm x$, \textbf{\emph{a$^k_n$}}) as $\bm 0$.
\STATE Select an initial state $\bm x$.
\FOR {each step}
\IF{$rand(.) < \epsilon$}
\STATE Randomly choose one action \textbf{\emph{a$^k_n$}} from \textbf{\emph{A$^k_n$}}.
\ELSE
\STATE Select the action $\textbf{\emph{a$^k_n$}} = \mathop {\arg \max }\limits_{\textbf{\emph{a$^{k'}_n$}}} {\emph{Q}}\left(\bm x', \textbf{\emph{a$^{k'}_n$}} \right)$.
\vspace{-5pt}
\ENDIF
\STATE Execute \textbf{\emph{a$^k_n$}}, obtain $R\left(\bm x, \bm a_n^k\right)$ and observe $\bm x'$.

\WHILE{$k<G \times B$}

\IF{$\bm x$ and \textbf{\emph{a$^k_n$}} have been recorded in the stacks}
\STATE Skip to step 4.
\ELSE
\STATE Record $\bm x$ and \textbf{\emph{a$^k_n$}} in the corresponding stack, $q=1$.
\ENDIF
\ENDWHILE
\STATE Execute $\bm a_n^k$, obtain $R\left(\bm x, \bm a_n^k\right)$ and observe $\bm x'$.
\STATE Update $Q\left(\bm x, \bm a^k_n\right)$ using (23), $\bm x=\bm x'$.
\ENDFOR
\end{algorithmic}
\end{algorithm}

\subsection{The Complexity of Multi-stack RL Method}
Next, we analyze the complexity of the proposed algorithm. Since the objective of the proposed algorithm is to find the optimal resource allocation policy, the complexity of the proposed algorithm depends on the number of actions in Q-table of each BS. Since the worst-case for each BS is to explore all actions, the worst-case complexity of the proposed algorithm is  $O(\left| \mathcal{A_{\rm 1}}\times \ldots \times \mathcal{A_{\rm n}} \right|)$, where $\left| \mathcal{A}_n\right|$ is the total number of actions of each BS \emph{n}. Thus, the number of actions of each BS \emph{n}, $\left| \mathcal{A}_n\right|$ can be given by the following theorem.
\begin{theorem}\label{thm1}{\rm Given the number of downlink and uplink subcarriers, $I$ and $J$, the number of the integer-valued transmit power levels, $N_a$, the number of actions per each BS \emph{n}, $\left| \mathcal{A_{\emph{n}}}\right|$, is given by:}
\end{theorem}
\begin{equation}\tag{24}
\vspace{-0.2cm}
\begin{aligned}
\left| {A_n^{}} \right| =& \sum\limits_{m \in \mathcal{M}} {\prod\limits_{i = 1}^{\left\| {\bm d}_{n,m} \right\|} {\left( {\begin{array}{*{20}{l}}
{\kern 20pt}{m_{i}}\\
{J - \sum\limits_{k = 1}^{i - 1} {m_{i}} }
\end{array}} \right)} } \times N_a^J
\times\!\!\! \sum\limits_{{m} \in \mathcal{M}} \!\!{\prod\limits_{i = 1}^{\left\|{\bm u}_{n,m}\right\|} {\left( {\begin{array}{*{20}{l}}
{\kern 20pt}{m_{i}}\\
{I - \sum\limits_{k = 1}^{i - 1} {m_{i}} }
\end{array}} \right)} }  \!\times\! N_a^I \!\times\! {{M}^{\left\| {{\mu _n}} \right\|}}\!,\
\end{aligned}
\end{equation}where $\left(\begin{array}{l}{x}\\{y}\end{array}\right)=\frac{x(x - 1)\ldots(x - y + 1)}{y(y - 1)\ldots1}$. ${\left\| {\bm d}_{n,m} \right\|}$ and ${\left\| {\bm u}_{n,m} \right\|}$ represent the number of downlink and uplink subcarriers that are allocated to the users associated with BS $n$, respectively.

\begin{IEEEproof}See Appendix D.
\end{IEEEproof}

From Theorem 3, we can see that, as the number of users and subcarriers as well as the integer-valued transmit power levels increases, the number of actions increases. As the number of actions increases, the worst-case complexity of the proposed algorithm increases. Based on Theorem 3, the worst-case complexity occurs as all BSs select their optimal probability policies after traversing all other actions and environment states. In consequence, the proposed algorithm will degenerate into $\epsilon$-greedy exploration. Therefore, the probability of appearance of the worst-case
scenario is ${\left( {1 - \frac{\varepsilon }{{\left| {{A_1}} \right|}}} \right)^{\left| {{A_1}} \right| - 1}} \times \ldots \times{\left( {1 - \frac{\varepsilon }{{\left| {{A_n}} \right|}}} \right)^{\left| {{A_n}} \right| - 1}}$.$\footnote{Based on (20), for each BS, the probability that the optimal action is not selected at each iteration is ${\left( {1 - \frac{\varepsilon }{{\left| {{A_n}} \right|}}} \right)}$ and hence, the probability that
the optimal action is selected at the last iteration is ${\left( {1 - \frac{\varepsilon }{{\left| {{A_n}} \right|}}} \right)^{\left| {{A_n}} \right| - 1}}$. Therefore, the probability of all BSs select their optimal action at last iteration is ${\left( {1 - \frac{\varepsilon }{{\left| {{A_1}} \right|}}} \right)^{\left| {{A_1}} \right| - 1}} \times \ldots \times{\left( {1 - \frac{\varepsilon }{{\left| {{A_n}} \right|}}} \right)^{\left| {{A_n}} \right| - 1}}$.}$ In addition, our proposed learning algorithm can use multiple stacks to control the tradeoff between exploitation and exploration. In particular, the length of stacks determines the probability of exploration. For example, as the length of stacks increases, the probability of exploration increases, and, hence, the number of iterations that the proposed algorithm needed to converge decreases. However, increasing the probability of exploration results in the decreases of the exploitation probability. In consequence, the BSs may not select the optimal action to decrease the transmission and computational delay. Moreover, the proof of convergence for the proposed algorithm, we can invoke our result in the fact that the algorithm will reach a stable strategy for each BS to minimize the maximal computation and transmission delay among all users follows directly from [30, Th. 2].

\vspace{-0.1cm}
\section{Simulation Results}
In our simulations, an MEC-based network area having a radius of 100 m is considered with \emph{N} = 3  uniformly distributed BSs and \emph{M} = 6 uniformly distributed users. The values of other parameters are defined in Table II. For comparison purposes, we consider a baseline that is the Q-learning algorithm in \cite{YFYR}. For this Q-learning algorithm, the states, the actions, and the reward function are set to the same states, actions as well as reward function defined in our proposed algorithm. At each iteration, this Q-learning algorithm will select an action based on the $\epsilon$-greedy mechanism and, then, uses a Q-table to record the states, actions, and the successful resource allocation policy resulting from the actions that the BSs have implemented. Note that, most of the simulation results that focus on the convergence time is used to show the proposed algorithm enables the BSs to rapidly adjust their resource allocation schemes as the data size of each computational task requested by each user varies. All statistical results are averaged over 5000 independent runs.

\begin{table}
\centering
\renewcommand\arraystretch{1}
\caption{Simulation Parameters \cite{MY}}
\setlength{\tabcolsep}{1.mm}{
\begin{tabular}{|c|c|c|c|}
\hline
\textbf{Parameter}&\textbf{Value}&\textbf{Parameter}&\textbf{Value}\\
\hline
\emph{N}& 3 &${N_a}$& 10 \\
\hline
\emph{M}& 6&$\delta$&2 \\
\hline
$\emph{I}$& 9 &$\sigma^2_N$& -95 dBm\\
\hline
$\emph{J}$& 9 & $\omega$& $\left[1000,2000\right]$\\
\hline
\emph{B}& 150 &$\omega_\emph{m}$& 1500\\
\hline
\emph{P}$_{\rm U}$& 0.5 W& $\lambda_\emph{m}$&$\left[100, 400\right]$ kbits\\
\hline
\emph{P}$_{\rm B}$& 1 W& \emph{F}&100 GHz\\
\hline
$\emph{W}$& 3 MHz &$\emph{f}_\emph{m}$&0.5 GHz\\
\hline
\end{tabular}}
\vspace{-0.3cm}
\end{table}

Fig. 2 shows how the number of iterations required to converge changes as the learning rate $\alpha$ varies. From this figure, we can see that, as $\alpha$ increases, the number of iterations needed to converge decreases. The main reason is that as $\alpha$ is close to 0, each BS learns little information from the new action. Fig. 2 also shows that the number of iterations needed to converge decreases more slowly as $\alpha$ continues to increase. This is due to the fact that as $\alpha$ is larger than 0.7, each BS has learned the information from the action and, hence, increasing $\alpha$ will not affect the convergence speed. From Fig. 2, we can also see that the proposed algorithm can reduce 25\% number of iterations needed to converge compared to the classical Q-learning algorithm. This is because the proposed algorithm enables the BSs to learn the historical resource allocation schemes and users' states that are recorded in stacks so as to increase the probability of exploration.

In Fig. 3, we show how the number of iterations required to converge varies as the value of discount factor $\gamma$ changes. Here, $\gamma = 0$ indicating that each BS emphasizes on the immediate reward, and $\gamma=1$ indicating that each BS emphasizes on the future reward. In this figure, we can see that, as $\gamma$ increases, the number of iterations needed for convergence increases at first and then decreases. This is due to the fact that as $\gamma$ decreases to 0, each BS only focuses on the immediate reward and chooses the optimal current action, which can control the errors resulting from an incorrect update of future steps. Hence, for each step, each BS can learn correct information so as to speed up the convergence. Meanwhile, as $\gamma$ increases to 1, each BS emphasizes on future reward, which means that each BS considers next actions for future steps and evaluates the current action, and hence, speeds up the convergence. Furthermore, as shown in Fig. 3, compared to Q-learning algorithm, the proposed method improves 18\% number of iterations needed to converge due to the fact that each BS learns environmental and users' information using multiple stacks.

\begin{figure}[t]
\centering
\setlength{\belowcaptionskip}{-0.4cm}
\includegraphics[width=9cm]{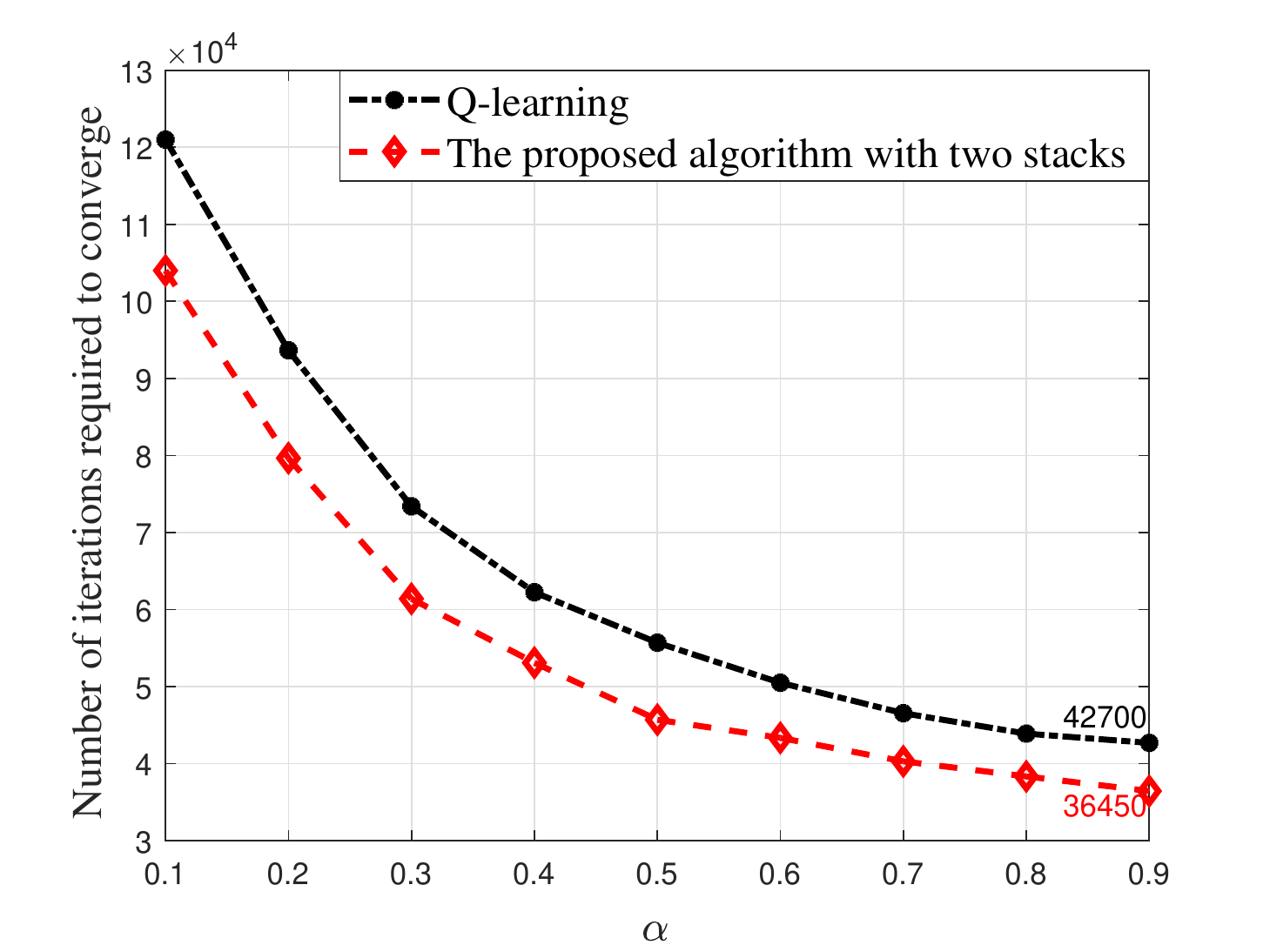}
\caption*{Fig. 2. Number of iterations required to converge as the learning rate $\alpha$ varies.}
\label{4}
\end{figure}
\begin{figure}[t]
\centering
\setlength{\belowcaptionskip}{-0.6cm}
\includegraphics[width=9cm]{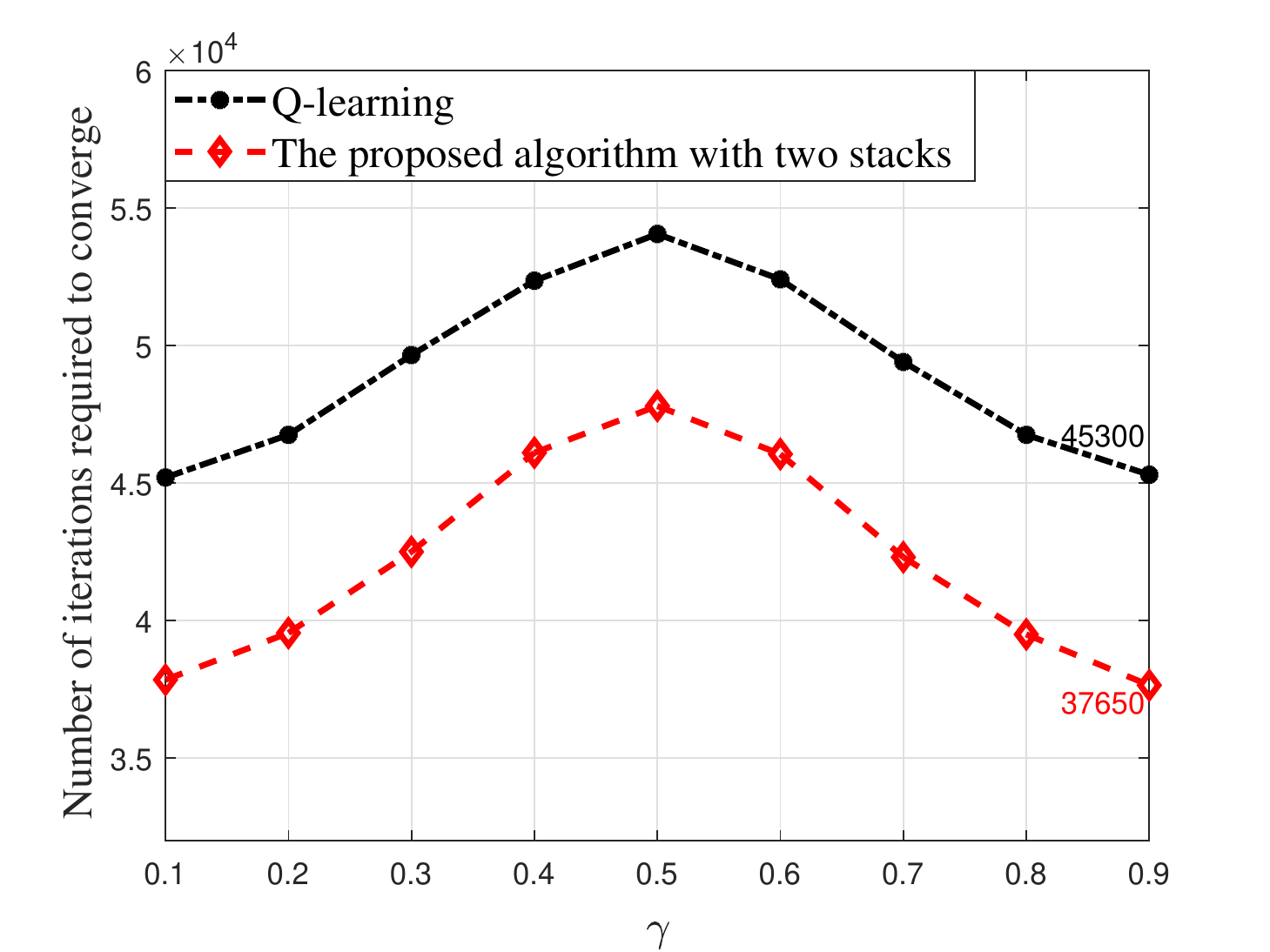}
\caption*{Fig. 3. Number of iterations required to converge as the discount factor $\gamma$ varies.}
\label{4}
\end{figure}

\begin{figure}[t]
\centering
\includegraphics[width=9cm]{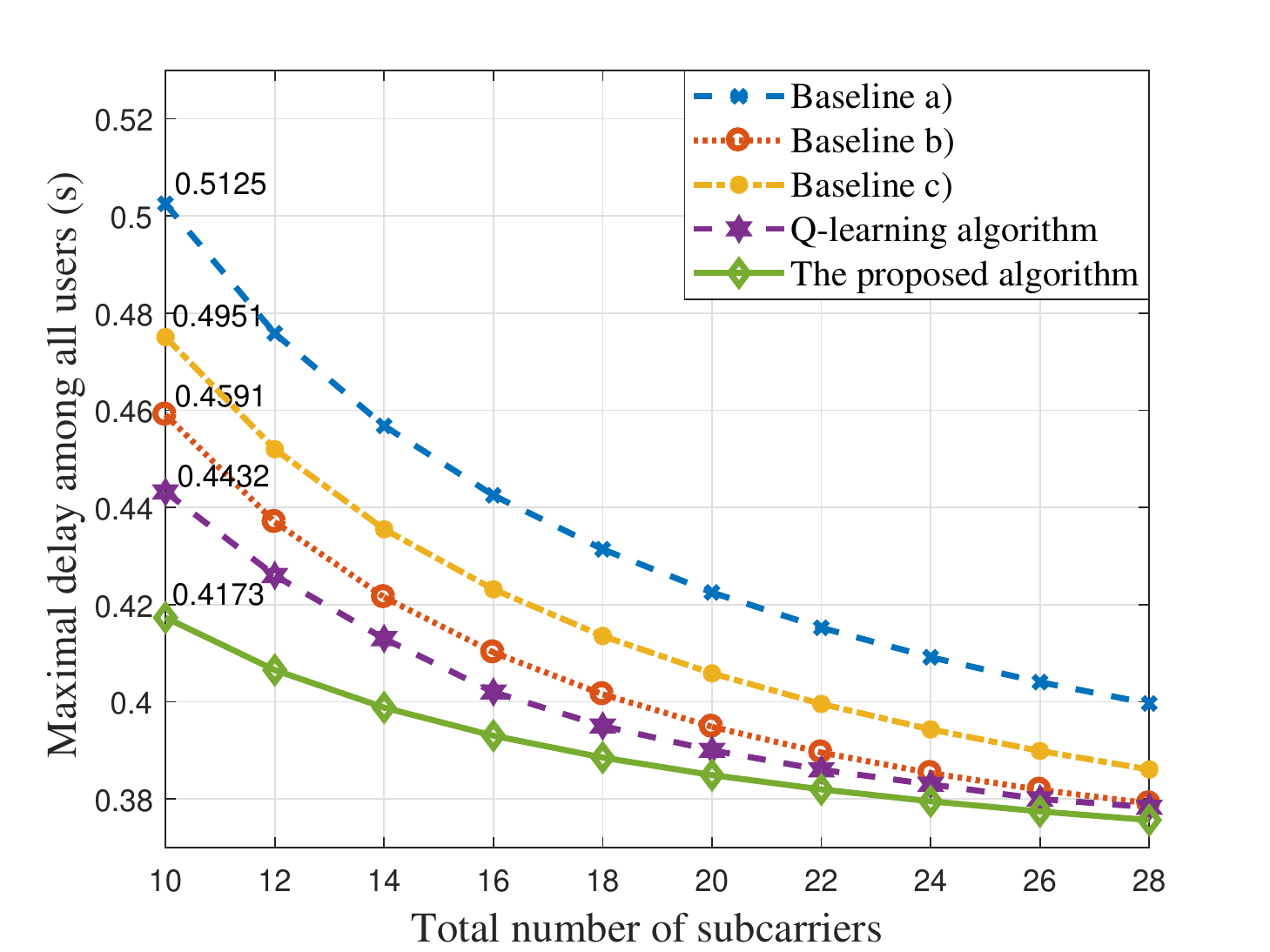}
\caption*{Fig. 4. Maximal delay changes as the total number of subcarriers $I$ and $J$ varies.}
\label{2}
\vspace{-0.5cm}
\end{figure}

In Fig. 4, we show how the maximal delay among all users \emph{T}$_{\rm max}$ changes as the number of subcarriers varies. In this figure, we consider three baselines:a) the optimization for task allocation with random subcarrier and power allocation, b) joint optimization of task and subcarrier allocation with random power allocation (i.e., $\bm a_{n}^k=[\bm u_{n},\bm d_{n}]$), and c) joint optimization for task and power allocation with random subcarrier allocation (i.e., $\bm a_{n}^k=[\bm v_{n},\bm w_{n}]$). Fig. 4 shows the maximal delay \emph{T}$_{\rm max}$ decreases as the number of subcarriers increases. The reason is that, as the number of subcarriers increases, the data rate between the BSs and each user increases, thus, the transmission delay decreases. From this figure, we can also see that the maximal delay decreases rapidly as the number of subcarriers is less than 20. As the number of subcarriers continues to increase, this decrease is limited. The reason is that, as the number of subcarriers is smaller than 20, the maximal delay of the users is decided by both computational and transmission delay. As the number of subcarriers continues to increase, the transmission delay is minimized thus, the maximal delay is dominated by the computational delay which is not minimized. From Fig. 4, we can also see that the proposed algorithm can achieve up to 8.7\% improvement in terms of maximal delay compared to the algorithms that do not consider power allocation. This gain stems from the fact that the proposed algorithm can optimize transmit power allocation. Fig. 4 also shows that the proposed algorithm can achieve up to 12.1\% improvement in terms of maximal delay compared to the algorithms that do not consider subcarrier allocation. This gain stems from the fact that the proposed algorithm can optimize the subcarrier allocation. In addition, the proposed algorithm can achieve up to 5.8\% improvement in terms of maximal delay compared to Q-learning algorithm as shown in Fig. 4. This is because the proposed algorithm can use multiple stacks to record the historical resource allocation schemes and users' information. Using the recorded information, each BS can avoid repeatedly learning the same resource allocation scheme so as to speed up the convergence.


In Fig. 5, we show how the maximal delay among all users \emph{T}$_{\rm max}$ changes as the data size of each computational task varies. Fig. 5 shows that the maximal delay \emph{T}$_{\rm max}$ increases as the data size of each task increases. This is because as the data size of each requested task increases, the time consumption for computation and transmission increases. From Fig. 5, we can also see that, as the average data size of each task is 600 kbits, the proposed scheme reduces the maximal delay by up to 84\% and 61\% compared to the cases in which each computational task is fully computed at user and fully computed at the MEC server, respectively. This is because that the proposed scheme jointly allocates the limited resources based on each user's need. From this figure, we can also see that, as the data size of the computational task is 600 kbits, the proposed algorithm can achieve up to 11.1\% gain in terms of maximal delay compared to Q-learning algorithm. This is due to the fact that each BS learns the information of historical resource allocation schemes and users' states recorded in multiple stacks, thus improving learning efficiency.

\begin{figure}[t]
\centering
\setlength{\belowcaptionskip}{-0.3cm}
\includegraphics[width=9cm]{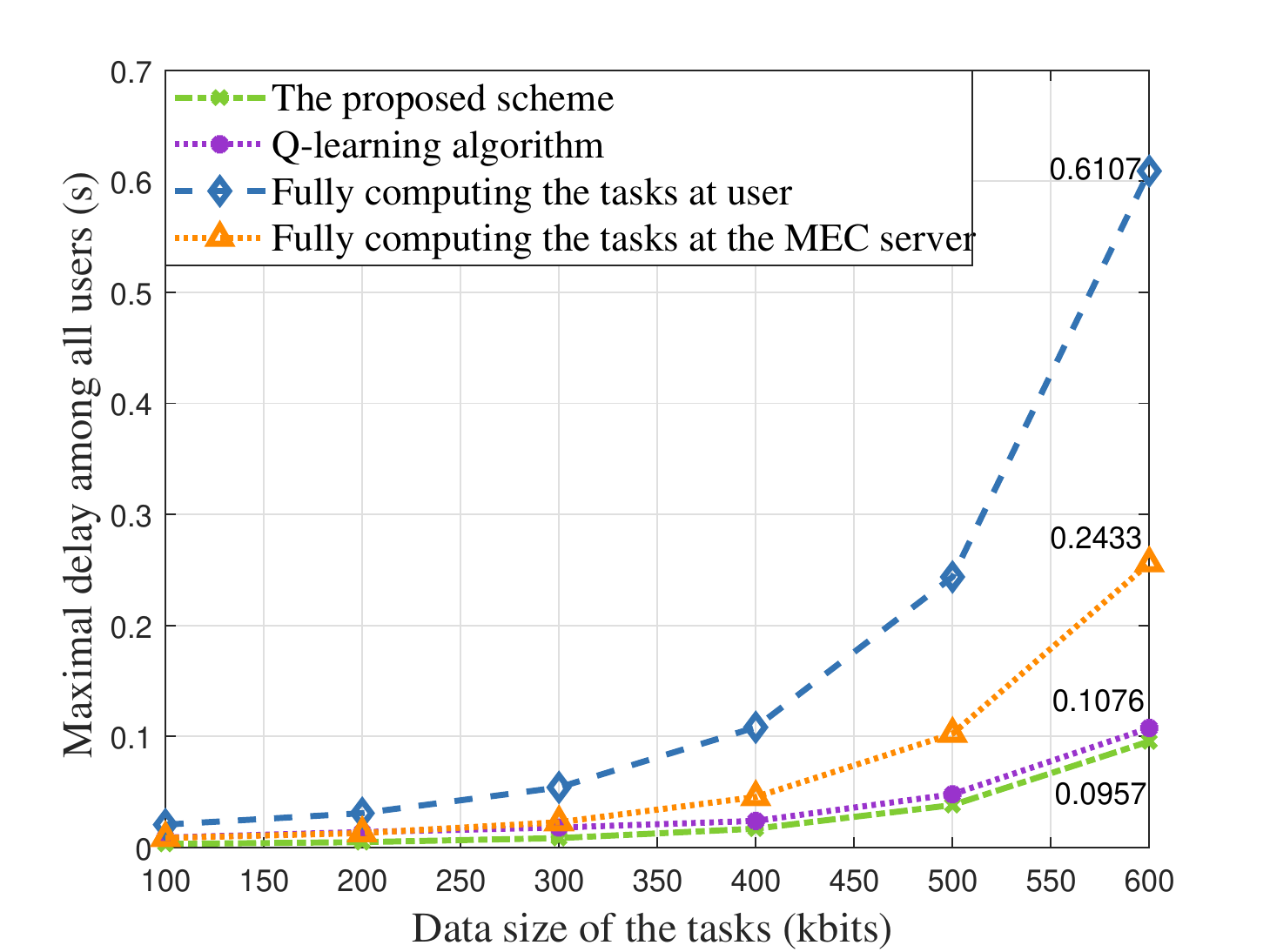}
\vspace{-0.1cm}
\caption*{Fig. 5. Maximal delay changes as the data size of tasks varies.}
\label{4}
\end{figure}

\begin{figure}
\centering
\includegraphics[width=9cm]{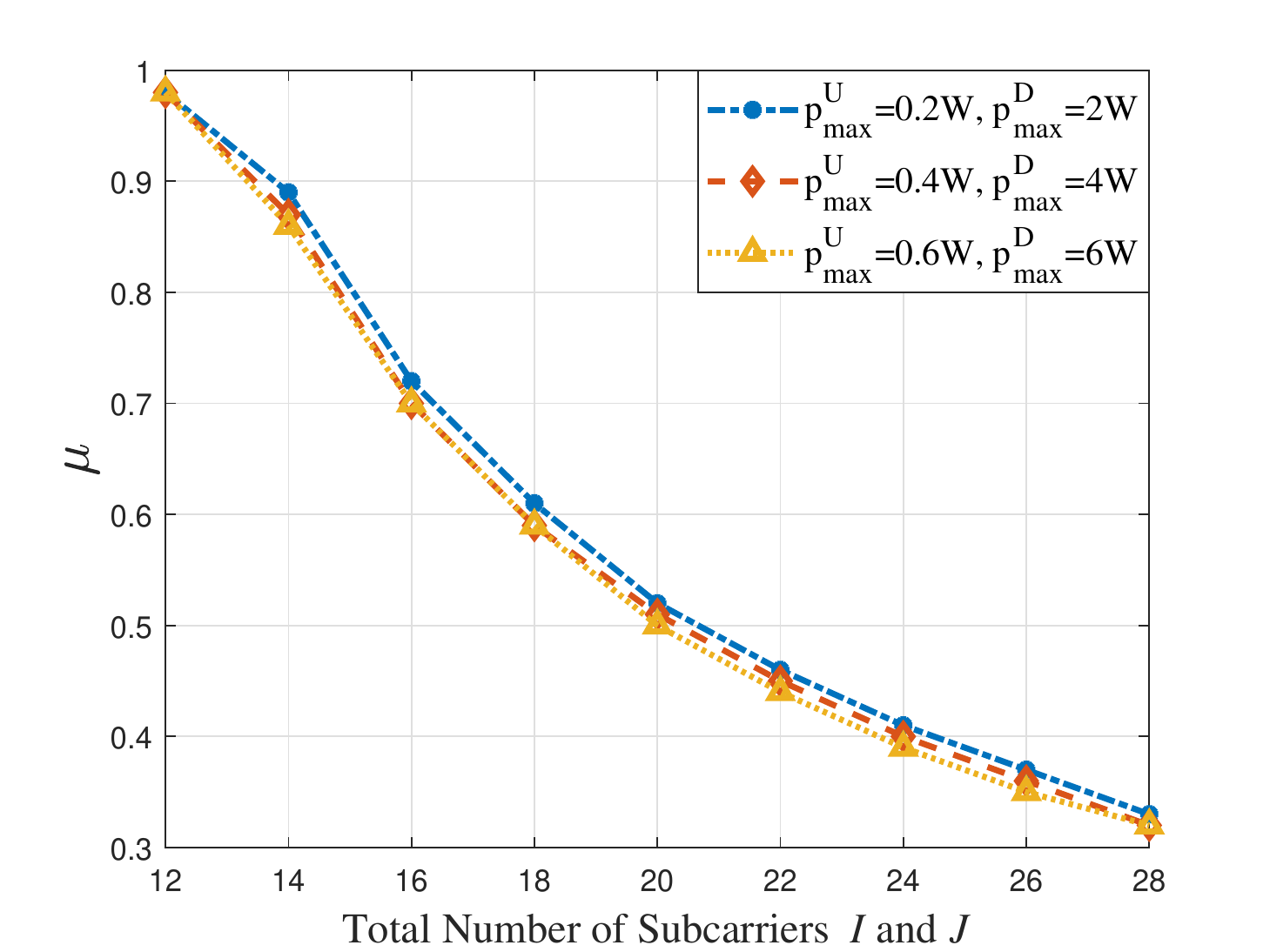}
\caption*{Fig. 6. Value of $\mu$ changes as the total number of subcarriers varies.}
\label{2}
\vspace{-0.3cm}
\end{figure}

Fig. 6 shows $\mu$ changes as the total number of subcarriers varies. From this figure, we can see that, as the total number of subcarriers and the transmit power of each BS increases, $\mu$ decreases. This is because as the total number of subcarriers increases, the data rate between the BSs and each user increases, and hence, the users can offload more computational tasks to the BSs that can use less time to compute the task than the users.

\begin{figure}[t]
\centering
\includegraphics[width=9cm]{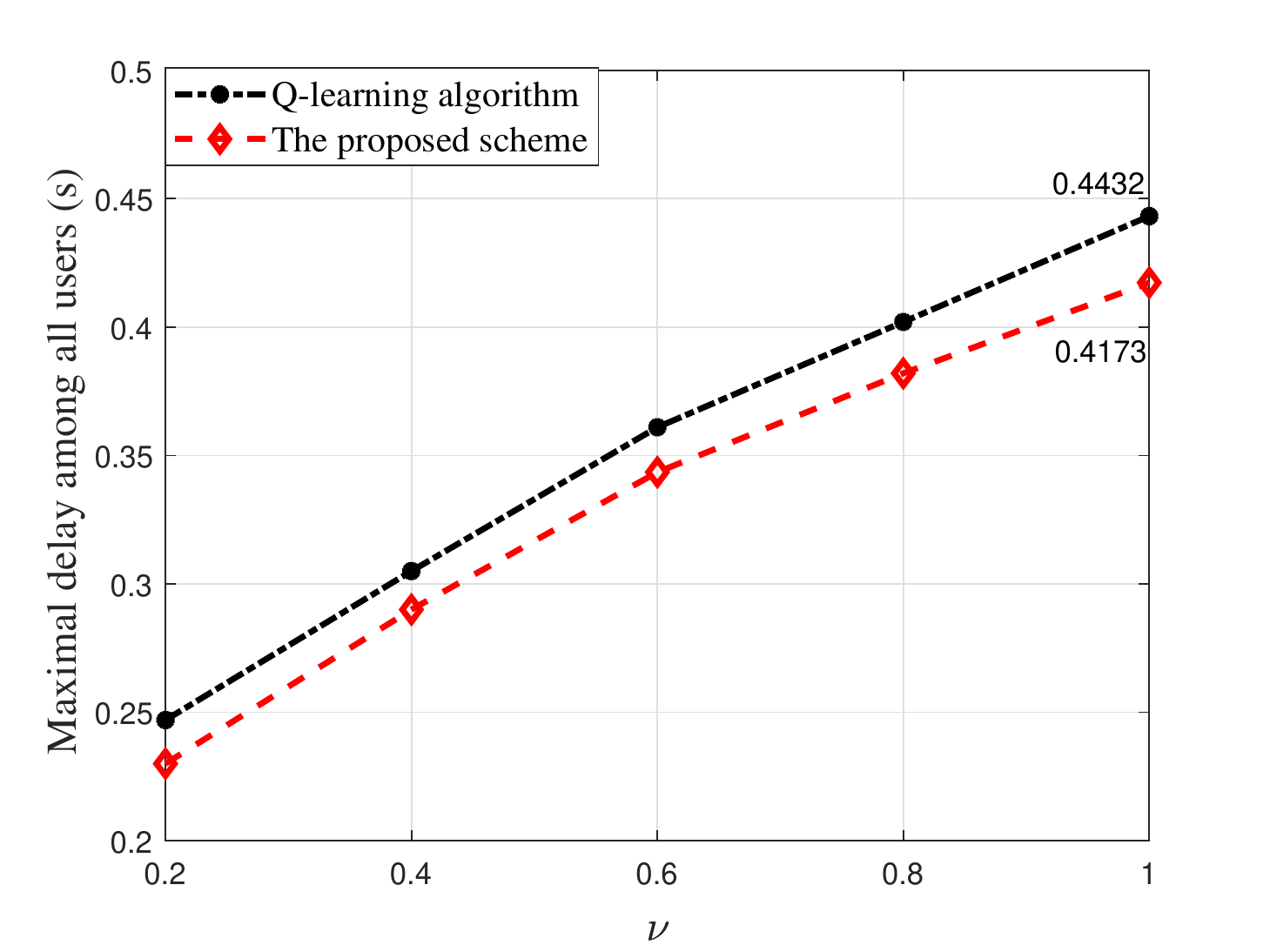}
\caption*{Fig. 7 Maximal delay changes as $\nu$ varies.}
\label{2}
\vspace{-0.55cm}
\end{figure}

Fig. 7 shows how $\nu$ affects the maximal delay among all users. From this figure, we can see that, as $\nu$ increases, the maximal delay among all users increases. The reason is that, as $\nu$ increases, the data size of the computational result of each computational task increases, and hence, the transmission delay increases. Fig. 7 also shows that the proposed algorithm can achieve up to 5.8\% gain in terms of maximal delay compared to Q-learning. This gain stems from the fact that the proposed algorithm enables each BS to avoid repeatedly learning the same resource allocation scheme so as to speed up the convergence.

\begin{figure}[t]
\centering
\vspace{-0.15cm}  
\includegraphics[width=9cm]{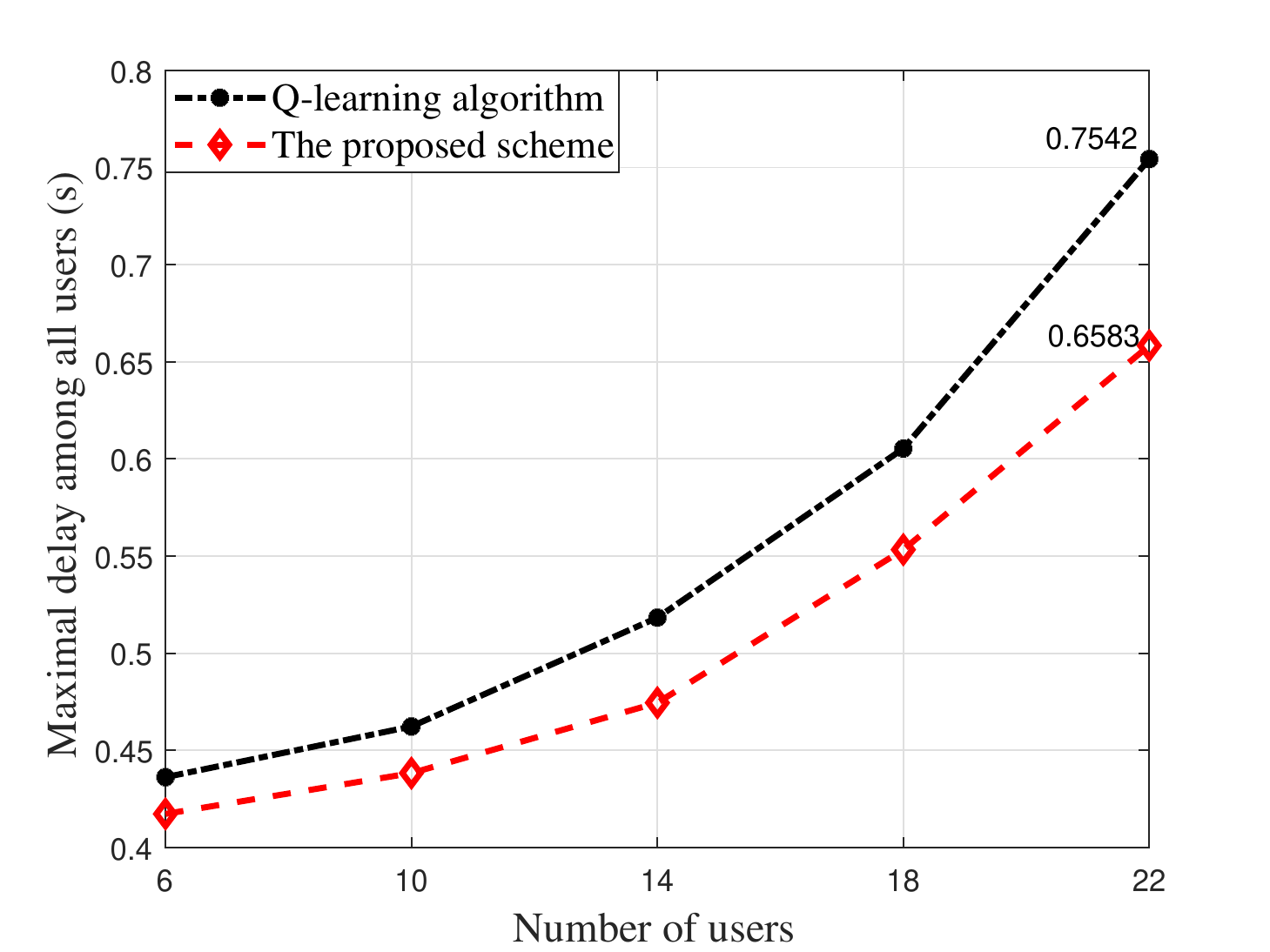}
\caption*{Fig. 8. Maximal delay changes as the number of users varies.}
\vspace{-0.3cm}
\label{2}
\end{figure}

Fig. 8 shows how the maximal delay changes as the number of users varies. From this figure, we can see that, the maximal delay among all users increases as the number of users increases. The reason is that as the number of users increases, the average number of subcarriers that can be allocated to each user decreases, and hence, the transmission delay increases. Fig. 8 also shows that the proposed algorithm can achieve up to 12.7\% gain in terms of maximal delay compared to Q-learning algorithm. This is because the proposed algorithm enables the BSs to record the historical resource allocation schemes and users' information so as to speed up the convergence and reduce the additional delay for computational task processing.

\section{Conclusion}
\vspace{-0.1cm}
In this paper, we have studied the problem of minimizing the maximal computation and transmission delay among all users that request diverse computational tasks. We have formulated the resource (subcarrier and transmit power) and task allocation problem as an optimization problem to meet the delay requirement of the users. A multiple stack RL method is proposed to solve this problem. Using the proposed algorithm, each BS records the historical resource allocation schemes and users' information in its multiple stacks that enable the BSs to record the historical resource allocation schemes and users' information in the stacks to improve learning efficiency and convergence speed. Simulation results show that the proposed algorithm can yields up to 18\% gain in terms of the number of iterations needed to converge compared to Q-learning algorithm. Meanwhile, the proposed scheme can achieve up to 11.1\% gain in terms of the maximal delay among all users compared to Q-learning algorithm.
\vspace{-0.1cm}
\section{Appendix}
\vspace{-0.1cm}
\subsection{Proof of Theorem 1}
\vspace{-0.1cm}
To prove Theorem 1, we first need to formulate the equation of the time used for processing the collaborative task, which is given by:
\begin{equation}\tag{7}
\begin{array}{l}
\begin{aligned}
&{t^3_m}\!\left( \! {\bm v_{n,m},\bm u_{n,m},\bm w_{n,m},\bm d_{n,m},{\mu _m}} \right)\\
&\!=\!\max \left( {\frac{{\omega _m}{\mu _m}{\lambda _m}}{f_m},}
\frac{{(1\!\!-\!{\mu _m}){\lambda _m}}}{U_{n,m}\!\!\left( {\bm v_{n,m},\!\bm u_{n,m}} \right)}{\rm{ + }}\frac{{\omega (1\!\!-\!{\mu _m}){\lambda _m}}}{F} \!+ \frac{{{\nu}(1\!\!-\!{\mu _m}){\lambda _m}}}{{{D_{n,m}}\!\!\left( {\bm{w}_{n,m},\!\bm d_{n,m}} \right)}}\!\right)\!\!,
\end{aligned}
\end{array}
\end{equation}
Obviously, as the time consumption for local computing ${\frac{{\omega _m}{\mu _m}{\lambda _m}}{f_m}}$ is equal to the time consumption for edge computing $\frac{{(1 - {\mu _m}){\lambda _m}}}{{{U_{n,m}}\left( {\bm v_{n,m},\bm u_{n,m}} \right)}}{\rm{ + }}\frac{{\omega (1 - {\mu _m}){\lambda _m}}}{F}
{ + \frac{{{\nu}(1 - {\mu _m}){\lambda _m}}}{{{D_{n,m}}\left( {\bm w_{n,m},\bm d_{n,m}} \right)}}}$, the minimum delay of a collaborative task is achieved, which can be expressed as:
\begin{equation}\tag{25}
\begin{aligned}
&\frac{{{\omega _m}{\mu _m}{\lambda _m}}}{{{f_m}}}=\frac{{(1\!-\!{\mu _m}){\lambda _m}}}{{{U_{n,m}}\!\!\left( {\bm{v}_{n,m},\!\bm{u}_{n,m}} \right)}}\!+\!\frac{{\omega (1 \!-\!{\mu _m}){\lambda _m}}}{F}\!+\!\frac{{(1\!-\!{\mu _m}){\lambda _m}\nu }}{{{D_{n,m}}\!\!\left( {\bm{w}_{n,m},\bm{d}_{n,m}} \right)}}.\
\end{aligned}
\end{equation}
Based on (25), the optimal $\mu_m$ can be given by:
\begin{equation}\tag{26}
\begin{aligned}
&{\mu _{m}} = \frac{\omega f_m+Y}{\omega f_m+Y+\omega_m F},\
\end{aligned}\label{7}
\end{equation}
where ${Y}{\rm{ = }}\frac{{f_m}F}{{U_{n,m}}(\bm{v}_{n,m},\bm{u}_{n,m})}+\frac{{f_m}\nu F}{{D_{n,m}}(\bm{w}_{n,m},\bm{d}_{n,m})}.$ This completes the proof.

\subsection{Proof of Theorem 2}

To capture the gain that stems from increasing the change of the number of the subcarriers and transmit power allocated to a user that has different computational tasks, we first need to change the number of downlink subcarriers $\bm{d}_{n,m}$, the number of uplink subcarriers $\bm{u}_{n,m}$, the downlink transmit power $\bm{w}_{n,m}$, and the uplink transmit power $\bm{v}_{n,m}$. Given the change of the allocated resource, the variation of the processing delay for each computational task can be given by:

For i), the gain that stems from increasing the number of the downlink subcarriers allocated to user \emph{m} that requests an edge task, $\Delta t^1_m (\bm{d}_{n,m}\!+\!\Delta \bm{d}_{n,m})$, is:
\begin{equation}\tag{27}
\begin{aligned}
\!\!&\Delta t^1_m (\bm{d}_{n,m}\!+\!\Delta \bm{d}_{n,m})\\
&=t_m^1\!\left( {\bm{w}_{n,m},\bm{d}_{n,m}} \right)\!-\!t_m^1\!\left( {\bm{w}_{n,m},\bm{d}_{n,m}\!+\!\Delta \bm d_{n,m}} \right) \\
&={\nu\lambda _m}\!\!\left(\!{\frac{1}{{{D_{n,m}}\!\!\left( {\bm{w}_{n,m},\bm{d}_{n,m}} \right)}}\!-\! \frac{1}{{{D_{n,m}}\!\!\left( {\bm{w}_{n,m},\bm{d}_{n,m}\!\!+\!\! \Delta \bm{d}_{n,m}} \right)}}}\!\right) \\
&= \frac{\nu{\lambda _m {D_{n,m}\!\left( {\bm{w}_{n,m},\Delta \bm{d}_{n,m}}\right)}}}{{{D_{n,m}}\!\left( {\bm{w}_{n,m},\!\bm{d}_{n,m}} \right){D_{n,m}}\!\left( {\bm{w}_{n,m},\!\bm{d}_{n,m}\!+\!\Delta \bm{d}_{n,m}} \right)}}\ \\
&=\frac{{\nu{\lambda _m}\Delta \bm{d}_{n,m}}}{{\bm{d}_{n,m}(\bm{d}_{n,m}\!+\!\Delta \bm{d}_{n,m}){D_{n,m}}\!\left( {\bm{w}_{n,m},\!\bm{d}_{n,m}} \right)}}.
\end{aligned}\label{7}
\end{equation}\vspace{-0.1cm}Here, when $\Delta\bm{d}_{n,m}\!\!\gg\!\!\bm{d}_{n,m}$, $\frac{\Delta\bm{d}_{n,m}}{\bm{d}_{n,m}(\bm{d}_{n,m}+\Delta\bm{d}_{n,m})}\!\!\approx\!\!\frac{1}{\bm d_{n,m}}$,~and, consequently, $\Delta t^1_m(\bm{d}_{n,m}\!+\!\Delta \bm{d}_{n,m})\!\!=\!\!\frac{{\nu{\lambda _m}}}{{\bm d_{n,m}{D_{n,m}}\left( {\bm{w}_{n,m},\bm{d}_{n,m}} \right)}}$. Moreover, as $\Delta\bm d_{n,m}$$\ll\bm d_{n,m}$, $\frac{\Delta\bm d_{n,m}}{\bm d_{n,m}(\bm d_{n,m}+\Delta\bm d_{n,m})}\!\!\approx\!\!\frac{{\Delta\bm d_{n,m}}}{(\bm d_{n,m})^2}$. Thus,  $\Delta t^1_m(\bm{d}_{n,m}\!+\!\Delta \bm{d}_{n,m})\!=\!\frac{{\nu{\lambda _m}\Delta \bm{d}_{n,m}}}{{\bm{d}_{n,m}^2{D_{n,m}}\left( {\bm{w}_{n,m},\bm{d}_{n,m}} \right)}}$.

For ii), the gain that stems from increasing the number of the uplink subcarriers allocated to user \emph{m} that requests a local task, $\Delta t^2_m(\bm{u}_{n,m}\!+\!\Delta \bm{u}_{n,m})$, is:
\begin{align}
&\Delta t^2_m(\bm{u}_{n,m}\!+\!\Delta \bm{u}_{n,m})\notag \\
&=\! t_m^2\left( {\bm{v}_{n,m},\bm{u}_{n,m} } \right) - t_m^2\left( {\bm{v}_{n,m},\bm{u}_{n,m}\!+\! \Delta \bm{u}_{n,m} }  \right)  \notag \\
&={\nu\lambda _m}\!\!\left(\!{\frac{1}{{{U_{n,m}}\!\left( {\bm{v}_{n,m},\!\bm{u}_{n,m}} \right)}}\!-\! \frac{1}{{{U_{n,m}}\!\left( {\bm{v}_{n,m},\!\bm{u}_{n,m}\!\!+\! \Delta \bm{u}_{n,m}} \right)}}}\!\right)\notag  \\
&=\! \frac{\nu{\lambda _m {U_{n,m}\!\left( {\bm{v}_{n,m},\Delta \bm{u}_{n,m}}\right)}}}{{{U_{n,m}}\left( {\bm{v}_{n,m},\!\bm{u}_{n,m}} \right){U_{n,m}}\left( {\bm{v}_{n,m},\!\bm{u}_{n,m}\!+\!\Delta \bm{u}_{n,m}} \right)}}\ \notag \\
&=\!\frac{{\nu{\lambda _m}\Delta \bm{u}_{n,m}}}{{\bm{u}_{n,m}(\bm{u}_{n,m}\!+\!\Delta \bm{u}_{n,m}){U_{n,m}}\left( {\bm{v}_{n,m},\!\bm{u}_{n,m}} \right)}} \tag{28}.
\end{align}\label{7}
\vspace{-0.5cm}

Similarly, when $\Delta\bm u_{n,m} \!\!\gg\!\! \bm u_{n,m}$, $\frac{\Delta\bm u_{n,m}}{\bm u_{n,m}(\!\bm u_{n,m}\!+\!\Delta\bm u_{n,m})} \!\!\approx\!\! \frac{1}{\bm u_{n,m}}$, and hence, $\Delta t^2_m(\bm{u}_{n,m}\!+\!\Delta \bm{u}_{n,m})\!\!=\!\!\frac{{\nu{\lambda _m}}}{\bm u_{n,m}{{U_{n,m}}\left( {\bm{v}_{n,m},\bm{u}_{n,m}} \right)}}$. Moreover, as $\Delta\bm u_{n,m} \!\!\ll\!\!\bm u_{n,m}$, $\frac{\Delta\bm u_{n,m}}{\bm u_{n,m}(\bm u_{n,m}+\Delta\bm u_{n,m})}\approx\frac{{\Delta\bm u_{n,m}}}{\bm u_{n,m}^2}$. Thus,  $\Delta t^2_m(\bm{u}_{n,m}\!+\!\Delta \bm{u}_{n,m})\!=\!\frac{{\nu{\lambda _m}\Delta \bm{u}_{n,m}}}{{\bm u_{n,m}^2{U_{n,m}}\left( {\bm{v}_{n,m},\bm{u}_{n,m}} \right)}}$.
\vspace{0.1cm}

For iii), since the CPU's performance of the MEC server is much better than that of the user's device, i.e., $\omega f_m \ll\omega_m F$, we have ${\mu _{m}} = \frac{\omega f_m+Y}{\omega f_m+Y+\omega_m F}\approx\frac{Y}{Y+\omega_m F}$. The gain that stems from increasing the number of the downlink subcarriers allocated to user \emph{m} that requests a collaborative task, $\Delta t^3_m(\bm{d}_{n,m}\!+\!\Delta \bm{d}_{n,m})$, is:
\vspace{-0.1cm}
\begin{align}
&\Delta t^3_m (\bm{d}_{n,m}\!+\!\Delta \bm{d}_{n,m}) \notag \\
&=\!\frac{{{\omega _m}{\lambda _m}{\mu _m}\!\!\left( {\bm{d}_{n,m},\!\bm{u}_{n,m}} \right)}}{{{f_m}}} {\rm{ - }}\frac{{{\omega _m}{\lambda _m}{\mu _m}\!\!\left( {\bm{d}_{n,m}\!\!+\!\Delta \bm{d}_{n,m},\!\bm{u}_{n,m}} \right)}}{{{f_m}}}\ \notag \\
&\approx \frac{{\omega _m}{\lambda _m}}{f_m}\!\!\left(\frac{Y}{Y\!+\!\omega_m F}\!-\!\frac{\Delta Y_d}{ \Delta Y_d\!+\!\omega_m F}\right) \notag \\
&={\lambda _m}\frac{\nu(\omega_m F)^2}{(\omega_m F+ Y)(\omega_m F+\Delta Y_d)} \times\!\!\left( {\frac{1}{{{D_{n,m}}\!\!\left( {\bm{w}_{n,m},\bm{d}_{n,m}} \right)}}\!-\! \frac{1}{{D_{n,m}}\!\!\left( {\bm{w}_{n,m},\bm{d}_{n,m}\!\!+\!\! \Delta \bm{d}_{n,m}} \right)}} \right) \notag \\
&=\!\frac{{\lambda _m}\nu(\omega_m F)^2\Delta \bm{d}_{n,m}}{{\bm{d}_{\!n,m}}(\!\bm{d}_{\!n,m}\!\!+\!\! \Delta \bm{d}_{n,m})(\omega_m F\!\!+\!\! Y)(\omega_m F\!\!+\!\!\Delta Y_d){D_{\!n,m}}\!\!\left(\! {\bm w_{n,m},\!\bm{d}_{n,m}}\! \right)} \tag{29}\!,
\end{align}\label{7}
where ${Y}=\frac{{f_m}F}{{U_{n,m}}(\bm{v}_{n,m},\bm{u}_{n,m})}+\frac{{f_m}\nu F}{{D_{n,m}}(\bm{w}_{n,m},\bm{d}_{n,m})}$ and $\Delta Y_d= \frac{{f_m}F}{{U_{n,m}}(\bm{v}_{n,m},\bm{u}_{n,m})}\!+\!\frac{{f_m}\nu F}{{D_{n,m}}(\bm{w}_{n,m},\bm{d}_{n,m}+\Delta\bm d_{n,m})}.$ Here, when $\Delta\bm d_{n,m} \!\gg\! \bm d_{n,m}$, $\frac{\Delta\bm d_{n,m}}{\bm d_{n,m}(\bm d_{n,m}\!+\!\Delta\bm d_{n,m})}\!\!\approx\!\!\frac{1}{\bm d_{n,m}}$, and, consequently, $\Delta t^3_m(\bm{d}_{n,m}\!\!+\!\!\Delta \bm{d}_{n,m})\!\!=\!\!\frac{\lambda _m\nu(\omega_m F)^2}{(\omega_m F+ Y)(\omega_m F+\Delta Y_d){D_{n,m}}\left( {\bm{w}_{n,m},\bm{d}_{n,m}} \right)}
$. Moreover, as $\Delta\bm d_{n,m}\!\! \ll \!\!\bm d_{n,m}$, \!$\frac{\Delta\bm d_{n,m}}{\bm d_{n,m}\!(\bm d_{n,m}\!+\!\Delta\bm d_{n,m})}\! \approx\!  \frac{{\Delta\bm d_{n,m}}}{\bm d_{n,m}^2}$. \!Thus, $\Delta t^3_m\!(\bm{d}_{n,m}\!+\!\Delta \bm{d}_{n,m})\!\!=\!\!
\frac{{{\lambda _m}\nu(\omega_m F)^2\Delta \bm{d}_{n,m}}}{{\bm{d}_{n,m}^2(\omega_m F\!+\! Y)(\omega_m F+\Delta Y_d){D_{n,m}}\!\left({\bm{w}_{n,m},\bm{d}_{n,m}} \right)}}$.

\vspace{0.05cm}
Similarly, the gain that stems from increasing the number of uplink subcarriers allocated to user \emph{m} that requests a collaborative task, $\Delta t^3_m(\bm{u}_{n,m}\!+\!\Delta \bm{u}_{n,m})$, is:
\begin{equation}\tag{30}
\begin{aligned}
&\Delta t^3_m (\bm{u}_{n,m}\!+\!\Delta \bm{u}_{n,m})\\
&=\! \frac{{{\omega _m}{\lambda _m}{\mu _m}\!\!\left(\! {\bm{d}_{n,m},\!\bm{u}_{n,m}} \right)}}{f_m} \!-\!\frac{{{\omega _m}{\lambda _m}{\mu _m}\!\!\left(\!{\bm{d}_{n,m} ,\!\bm{u}_{n,m}}\!\!+\!\!\Delta \bm{u}_{n,m} \right)}}{{{f_m}}}\ \\
&\approx \frac{{\omega _m}{\lambda _m}}{f_m}\!\!\left(\frac{Y}{Y\!+\!\omega_m F}\!-\!\frac{\Delta Y_u}{ \Delta Y_u\!+\!\omega_m F}\right) \\
&=\! \frac{\omega_m\lambda_m}{f_m}\!\frac{\omega_m F (Y\!-\!\Delta Y_u)}{(\omega_m F\!+\! Y)(\omega_m F\!+\!\Delta Y_u)} \\
&={\lambda _m}\!\frac{(\omega_m F)^2}{(\omega_m F\!\!+\!Y)(\omega_m F\!\!+\!\Delta Y_u)}
\times\!\!\left( {\frac{1}{{{U_{n,m}}\!\!\left( {\bm{v}_{n,m},\bm{u}_{n,m}} \right)}}\!-\! \frac{1}{{{U_{n,m}}\!\!\left( {\bm{v}_{n,m},\bm{u}_{n,m}\!\!+\!\! \Delta \bm{u}_{n,m}} \right)}}} \right) \\
&=\!\frac{{{\lambda _m}(\!\omega_m F)^2\Delta \bm{u}_{n,m}}}{{\bm{u}_{\!n,m}\!(\!\bm{u}_{n,m}\!\!+\!\! \Delta \bm{u}_{n,m}\!)(\!\omega_m\!F\!\!+\!\! Y\!)(\omega_m\!F\!\!+\!\!\Delta Y_u\!){U_{\!n,m}}\!\!\left( {\bm{v}_{n,m},\bm{u}_{n,m}}\!\right)}}\!,
\end{aligned}\label{7}
\end{equation}
where ${Y}{\rm{ = }}\frac{{f_m}F}{{U_{n,m}}(\bm{v}_{n,m},\bm{u}_{n,m})}+\frac{{f_m}\nu F}{{D_{n,m}}(\bm{w}_{n,m},\bm{d}_{n,m})}$ and ${\Delta Y_u}=\frac{{f_m}F}{{U_{n,m}}(\bm{v}_{n,m},\bm{u}_{n,m}+\Delta\bm u_{n,m})}+\frac{{f_m}\nu F}{{D_{n,m}}(\bm{w}_{n,m},\bm{d}_{n,m})}.$ Here, when $\Delta\bm u_{n,m} \!\!\gg\!\!\bm u_{n,m}$, $\frac{\Delta\bm u_{n,m}}{\bm u_{n,m}(\bm u_{n,m}+\Delta\bm u_{n,m})}\!\!\approx\!\!\frac{1}{\bm u_{n,m}}$, and, consequently, $\Delta t^3_m(\bm{u}_{n,m}\!+\!\Delta \bm{u}_{n,m}) = \frac{\lambda _m(\omega_m F)^2}{(\omega_m F+ Y)(\omega_m F+\Delta Y_d){U_{n,m}}\left( {\bm{v}_{n,m},\bm{u}_{n,m}} \right)}$.
Moreover, as $\Delta\bm u_{n,m} \!\!\ll\!\!\bm u_{n,m}$, \!$\frac{\Delta\bm u_{n,m}}{\bm u_{n,m}(\bm u_{n,m}\!+\!\Delta\bm u_{n,m})}\!\!\!\approx\!\!\frac{{\Delta\bm u_{n,m}}}{\bm u_{n,m}^2}$. \!Thus, $\Delta t^3_m\!(\!\bm{u}_{n,m}\!+\!\Delta \bm{u}_{n,m})\!=\!\frac{{\lambda _m}(\omega_m F)^2\Delta \bm{u}_{n,m}}{(\!\omega_m F\!+\! Y\!)(\!\omega_m F\!+\!\Delta Y_u\!){\bm u_{n,m}^2{U_{n,m}}\!\left( {\bm{v}_{n,m},\bm{u}_{n,m}} \right)}}$.

\vspace{0.05cm}
For iv), the gain that stems from increasing the transmit power allocated to user \emph{m} that requests an edge task, $\Delta t^1_m(\bm{w}_{n,m}\!+\!\Delta \bm{w}_{n,m})$, is:
\begin{equation}\tag{31}
\begin{aligned}
\Delta t^1_m (\bm{w}_{n,m}\!+\!\Delta \bm{w}_{n,m})\!&=\!t_m^1\!\!\left( {\bm{w}_{n,m},\!\bm{d}_{n,m}} \right)\!-\!t_m^1\left( {\bm{w}_{n,m}\!\!+\!\Delta \bm{w}_{n,m},\!\bm{d}_{n,m}} \right) \\
&=\! \frac{\nu{\lambda _m}({{D_{n,m}}\!\left( {\bm{w}_{n,m}\!\!+\!\!\Delta \bm{w}_{n,m},\!\bm{d}_{n,m}} \right)}\!-\!{{D_{n,m}}\!\left( {\bm{w}_{n,m},\!\bm{d}_{n,m}} \right))}}{{{D_{n,m}}\!\left( {\bm{w}_{n,m},\!\bm{d}_{n,m}} \right)}{{D_{n,m}}\!\left( {\bm{w}_{n,m}\!\!+\!\!\Delta \bm{w}_{n,m},\!\bm{d}_{n,m}} \right)}}.\
\end{aligned}\label{7}
\end{equation}

For v), the gain that stems from increasing the transmit power allocated to user \emph{m} that requests a local task, $\Delta t^2_m(\bm{v}_{n,m}\!+\!\Delta \bm{v}_{n,m})$, is:
\begin{equation}\tag{32}
\begin{aligned}
\Delta t^2_m (\bm{v}_{n,m}\!+\!\Delta \bm{v}_{n,m})&=\!t_m^2\left( {\bm{v}_{n,m},\bm{u}_{n,m}} \right) \!-\! t_m^2\left( {\bm{v}_{n,m}\! +\! \Delta \bm{v}_{n,m},\bm{u}_{n,m}} \right) \\
&=\! \frac{\nu{\lambda _m}({{U_{n,m}}\!\left( {\bm{v}_{n,m}\!\!+\!\!\Delta \bm{v}_{n,m},\!\bm{u}_{n,m}} \right)}\!-\!{{U_{n,m}}\!\left( {\bm{v}_{n,m},\!\bm{u}_{n,m}} \right))}}{{{U_{n,m}}\!\left( {\bm{v}_{n,m},\!\bm{u}_{n,m}} \right)}{{U_{n,m}}\!\left( {\bm{v}_{n,m}\!\!+\!\!\Delta \bm{v}_{n,m},\!\bm{u}_{n,m}} \right)}}\! .\
\end{aligned}\label{7}
\end{equation}

For vi), the gain that stems from increasing the transmit power on uplink subcarriers allocated to user \emph{m} that requests a collaborative task, $\Delta t^3_m(\bm{v}_{n,m}\!+\!\Delta \bm{v}_{n,m})$, is:
\begin{align}
&\Delta t^3_m(\bm{v}_{n,m}\!+\!\Delta \bm{v}_{n,m}) \notag\\
&=\!\frac{{{\omega _m}{\lambda _m}{\mu _m}\!\!\left( {\bm{w}_{n,m},\!\bm{v}_{n,m}} \right)}}{{{f_m}}} {\rm{ - }}\frac{{{\omega _m}{\lambda _m}{\mu _m}\!\!\left( {\bm{w}_{n,m} ,\!\bm{v}_{n,m}\!\!+\!\!\Delta \bm{v}_{n,m} } \right)}}{{{f_m}}}\ \notag\\
&\approx  \frac{{\omega _m}{\lambda _m}}{f_m}\!\!\left(\frac{Y}{Y\!+\!\omega_m F}\!-\!\frac{\Delta Y_v}{ \Delta Y_v\!+\!\omega_m F}\right) \notag\\
&=\!\frac{{\lambda _m}(\omega_m F)^2}{(\omega_m F\!+\!Y)(\omega_m F\!+\!\Delta Y_v)}\times\!\!\frac{{{U_{n,m}}\!\left( {\bm{v}_{n,m}\!\!+\!\!\Delta \bm{v}_{n,m},\!\bm{u}_{n,m}} \right)}\!-\!{{U_{n,m}}\!\left( {\bm{v}_{n,m},\!\bm{u}_{n,m}} \right)}}{{{U_{n,m}}\!\left( {\bm{v}_{n,m},\!\bm{u}_{n,m}} \right)}{{U_{n,m}}\!\left( {\bm{v}_{n,m}\!\!+\!\!\Delta \bm{v}_{n,m},\!\bm{u}_{n,m}} \right)}}\!,\tag{33}
\end{align}\label{7}
where ${Y}{\rm{ = }}\frac{{f_m}F}{{U_{n,m}}(\bm{v}_{n,m},\bm{u}_{n,m})}+\frac{{f_m}\nu F}{{D_{n,m}}(\bm{w}_{n,m},\bm{d}_{n,m})}$ and ${\Delta Y_v}=\frac{{f_m}F}{{U_{n,m}}(\bm{v}_{n,m}+\Delta\bm v_{n,m},\bm{u}_{n,m})}+\frac{{f_m}\nu F}{{D_{n,m}}(\bm{w}_{n,m},\bm{d}_{n,m})}.$

\vspace{0.1cm}
Similarly, the gain that stems from increasing the transmit power on downlink subcarriers allocated to user \emph{m} that requests a collaborative task, $\Delta t^3_m(\bm{w}_{n,m}\!+\!\Delta \bm{w}_{n,m})$, is:
\begin{equation}\tag{34}
\begin{aligned}
&\Delta t^3_m(\bm{v}_{n,m}\!+\!\Delta \bm{v}_{n,m})\\
&=\!\frac{{{\omega _m}{\lambda _m}{\mu _m}\!\!\left( {\bm{w}_{n,m}\!,\!\bm{v}_{n,m}} \right)}}{{{f_m}}} {\rm{ - }}\frac{{{\omega _m}{\lambda _m}{\mu _m}\!\!\left( {\bm{w}_{n,m}\!\!+\!\! \Delta \bm w_{n,m}\!,\!\bm{v}_{n,m} } \right)}}{{{f_m}}}\ \\
&\approx  \frac{{\omega _m}{\lambda _m}}{f_m}\!\!\left(\frac{Y}{Y\!+\!\omega_m F}\!-\!\frac{\Delta Y_w}{ \Delta Y_w\!+\!\omega_m F}\right) \\
&=\!\frac{{\lambda _m}\nu(\omega_m F)^2}{(\omega_m F\!+\! Y)(\omega_m F\!+\!\Delta Y_w)}
\times \!\frac{{{D_{n,m}}\!\left( {\bm{w}_{n,m}\!\!+\!\!\Delta \bm{w}_{n,m},\!\bm{d}_{n,m}} \right)}\!-\!{{D_{n,m}}\!\left( {\bm{w}_{n,m},\!\bm{d}_{n,m}} \right)}}{{{D_{n,m}}\!\left( {\bm{w}_{n,m},\!\bm{d}_{n,m}} \right)}{{D_{n,m}}\!\left( {\bm{w}_{n,m}\!\!+\!\!\Delta \bm{w}_{n,m},\!\bm{d}_{n,m}} \right)}}\!,
\end{aligned}\label{7}
\end{equation}
where ${Y}{\rm{ = }}\frac{{f_m}F}{{U_{n,m}}(\bm{v}_{n,m},\bm{u}_{n,m})}+\frac{{f_m}\nu F}{{D_{n,m}}(\bm{w}_{n,m},\bm{d}_{n,m})}$ and ${\Delta Y_w}=\frac{{f_m}F}{{U_{n,m}}(\bm{v}_{n,m},\bm{u}_{n,m})}+\frac{{f_m}\nu F}{{D_{n,m}}(\bm{w}_{n,m}+\Delta\bm w_{n,m},\bm{d}_{n,m})}.$

\vspace{0.1cm}
This completes the proof.
\subsection{Proof of Collary 1}
To find the relationship among the gains that stem from the change of the same number of subcarriers or transmit power for a user that has different computational tasks, we first need to prove that for a user that requests an edge task, increasing the number of uplink subcarriers or uplink transmit power will not change the delay. From (5), we can see that the delay of a user that requests an edge task depends on the downlink subcarriers and downlink transmit power. In consequence, increasing the number of uplink subcarriers or transmit power will not affect the downlink transmission rate, $\Delta t_m^{1}(\bm{u}_{n,m}+\Delta \bm{u}_{n,m})=\Delta t_m^{1}(\bm{v}_{n,m}+\Delta \bm{v}_{n,m})=0$. Similarly, from (6), we can see that increasing the number of downlink subcarriers or downlink transmit power of a user that requests a local task will not change the uplink transmission rate, which results in  $\Delta t_m^{2}(\bm{d}_{n,m}+\Delta \bm{d}_{n,m})=\Delta t_m^{2}(\bm{w}_{n,m}+\Delta \bm{w}_{n,m})=0$.

To find the relationship among the gains that stem from the change of the number of downlink subcarriers allocated to a user that has different computational tasks, we need to compare the delay gain of a user that requests an edge task as shown in (27) with the delay gain of a user that requests a collaborative task as shown in (29). In (29), since $\nu<1$ and hence, $\frac{\nu(\omega_m F)^2}{(\omega_m F+ Y)(\omega_m F+\Delta Y)}<1$, we have $\Delta t_m^{3}(\bm{d}_{n,m}+\Delta \bm{d}_{n,m})<\Delta t_m^{1}(\bm{d}_{n,m}+\Delta \bm{d}_{n,m})$. Then, based on $\Delta t_m^{2}(\bm{d}_{n,m}+\Delta \bm{d}_{n,m})=0$, we can obtain that $\Delta t_m^{2}(\bm{d}_{n,m}+\Delta \bm{d}_{n,m})<\Delta t_m^{3}(\bm{d}_{n,m}+\Delta \bm{d}_{n,m})<\Delta t_m^{1}(\bm{d}_{n,m}+\Delta \bm{d}_{n,m})$.

\vspace{0.1cm}
To analyze the gains that result from the change of the number of uplink subcarriers allocated to a user with different computational tasks, we need to compare the delay gain of a user that requests a local task as shown in (28) with the delay gain of a user that requests a collaborative task as shown in (30). In (30), since $\frac{(\omega_m F)^2}{(\omega_m F+ Y)(\omega_m F+\Delta Y)}<1$, we have $\Delta t_m^{3}(\bm{u}_{n,m}+\Delta \bm{u}_{n,m})<\Delta t_m^{2}(\bm{u}_{n,m}+\Delta \bm{u}_{n,m})$. Then, based on $\Delta t_m^{1}(\bm{u}_{n,m}+\Delta \bm{u}_{n,m})=0$, we can obtain that $\Delta t_m^{1}(\bm{u}_{n,m}+\Delta \bm{u}_{n,m})<\Delta t_m^{3}(\bm{u}_{n,m}+\Delta \bm{u}_{n,m})<\Delta t_m^{2}(\bm{u}_{n,m}+\Delta \bm{u}_{n,m})$.

\vspace{0.1cm}
To find the relationship among the gains that stem from the change of the downlink transmit power allocated to a user that has different computational tasks, we need to compare the delay gain of a user that requests an edge task as shown in (31) with the delay gain of a user that requests a collaborative task as shown in (34). In (34), since $\nu<1$, and hence, $\frac{\nu(\omega_m F)^2}{(\omega_m F+ Y)(\omega_m F+\Delta Y)}<1$, we have $\Delta t_m^{3}(\bm{w}_{n,m}+\Delta \bm{w}_{n,m})<\Delta t_m^{1}(\bm{w}_{n,m}+\Delta \bm{w}_{n,m})$. Then, based on $\Delta t_m^{2}(\bm{w}_{n,m}+\Delta \bm{w}_{n,m})=0$, we can obtain that $\Delta t_m^{2}(\bm{w}_{n,m}+\Delta \bm{w}_{n,m})<\Delta t_m^{3}(\bm{w}_{n,m}+\Delta \bm{w}_{n,m})<\Delta t_m^{1}(\bm{w}_{n,m}+\Delta \bm{w}_{n,m})$.

To analyze the gains that result from the change of the uplink transmit power allocated to a user with different computational tasks, we need to compare the delay gain of a user that requests a local task as shown in (32) with the delay gain of a user that requests a collaborative task as shown in (33). In (33), since $\frac{(\omega_m F)^2}{(\omega_m F+ Y)(\omega_m F+\Delta Y)}$
$<1$, we have $\Delta t_m^{3}(\bm{v}_{n,m}+\Delta \bm{v}_{n,m})<\Delta t_m^{2}(\bm{v}_{n,m}+\Delta \bm{v}_{n,m})$. Then, based on $\Delta t_m^{1}(\bm{v}_{n,m}+\Delta \bm{v}_{n,m})=0$, we can obtain that $\Delta t_m^{1}(\bm{v}_{n,m}+\Delta \bm{v}_{n,m})<\Delta t_m^{3}(\bm{v}_{n,m}+\Delta \bm{v}_{n,m})<\Delta t_m^{2}(\bm{v}_{n,m}+\Delta \bm{v}_{n,m})$.

{\vspace{0.1cm}}
This completes the proof.
\subsection{Proof of Theorem 3}
{\vspace{-0.1cm}}
To prove Theorem 3, we first need to derive the number of actions of each BS over the downlink subcarriers. Since each BS will allocate all downlink subcarriers to its associated users, for the first step, we assume that each BS allocates \emph{m$_1$} downlink subcarriers to the first user, and each BS has $\left(\begin{array}{l}
m_{1}\\J\end{array}\right)$ actions to allocate the downlink subcarriers to the first user. Based on the downlink subcarriers allocated to the first user, each BS allocates \emph{m$_2$} downlink subcarriers to the second user, each BS will have $\left( \begin{array}{l}{\quad \! m}_{2}\\J - m_{1}\end{array} \right)$ actions to allocate the downlink subcarriers to the second user. Using the enumeration method, the number of actions of each BS for downlink subcarrier allocation is $\sum\limits_{m \in \mathcal{M}} {\prod\limits_{i = 1}^{\left\| {\bm d}_{n,m} \right\|} {\left( \begin{array}{l}{\kern 20pt}m_{i}\\J - \sum\limits_{k = 1}^{i - 1} {m_k}\end{array} \right)}}$. Next, we formulate the number of actions for each BS for downlink transmit power allocation is $N_{a}^J$. Since each BS can allocate $J$ subcarriers to the users, and the number of transmit power actions on each subcarrier is $N_{a}$, the number of actions of each BS for transmit power allocation is $N_{a}^J$. In consequence, for the downlink, the total number of actions of each BS for subcarrier allocation and power allocation can be given by:
\begin{equation}\notag
\begin{aligned}
 \sum\limits_{m \in \mathcal{M}} {\prod\limits_{i = 1}^{\left\| {\bm d}_{n,m} \right\|} {\left( \begin{array}{l}{\kern 20pt}m_{i}\\J - \sum\limits_{k = 1}^{i - 1} {m_k}\end{array} \right)} } \times {N_{a}^J}.
\end{aligned}
\end{equation}
\quad The deviation of the number of actions of each BS for subcarrier and power allocation over the uplink is similar to the deviation of the number of actions over downlink, which is given by:
\begin{equation}\notag
\begin{aligned}
\sum\limits_{m \in \mathcal{M}} {\prod\limits_{i = 1}^{\left\| {\bm u}_{n,m} \right\|} {\left( \begin{array}{l}
{\kern 20pt}m{_i}\\
I - \sum\limits_{k = 1}^{i - 1} {m_k}
\end{array} \right)} }  \times {N_{a}^I}.
\end{aligned}
\end{equation}

In consequence, the number of actions per each BS is given by:
\begin{equation}\notag
\begin{aligned}
\!\!\sum\limits_{m \in \mathcal{M}}\!\! {\prod\limits_{i = 1}^{\left| {\bm d}_{n,m} \right|} \!\!{\left(\!\!\!\!\begin{array}{l}{\kern 20pt}m_{i}\\J - \sum\limits_{k = 1}^{i - 1} {m_k}\end{array}\!\!\!\!\right)}}\!\times\!N_{a}^J \times\!\! \sum\limits_{m \in \mathcal{M}}\!{\prod\limits_{i = 1}^{\left\| {\bm u}_{n,m} \right\|} \!\!{\left(\!\!\!\! \begin{array}{l}
{\kern 20pt}m{_i}\\
I - \sum\limits_{k = 1}^{i - 1} {m_k}
\end{array} \!\!\!\!\right)} }\!\!\times\!\! N_{a}^I\!.
\end{aligned}
\end{equation}

This completes the proof.

\renewcommand{\baselinestretch}{1.45}

\end{document}